\documentclass{egpubl}
\usepackage{eg2026}
 
\ConferencePaper        
\CGFStandardLicense

\usepackage[T1]{fontenc}
\usepackage{dfadobe}  

\BibtexOrBiblatex
\electronicVersion
\PrintedOrElectronic

\ifpdf \usepackage[pdftex]{graphicx} \pdfcompresslevel=9
\else \usepackage[dvips]{graphicx} \fi

\usepackage{mathtools}
\usepackage{amsfonts}
\usepackage{graphicx} 
\usepackage[nameinlink,capitalize]{cleveref}
\usepackage[ruled,lined,linesnumbered,procnumbered]{algorithm2e}
\usepackage{hyperref}
\usepackage{booktabs}
\usepackage[table]{xcolor}
\usepackage{subcaption}

\usepackage{float}
\usepackage{afterpage}
\usepackage{multirow}
\usepackage{dblfloatfix}
\usepackage{listings}
\usepackage{egweblnk} 

\title[HeatMat: Simulation of City Material Impact on UHI Effect]%
{HeatMat: Simulation of City Material Impact on\\Urban Heat Island Effect}

\author[M. Reinbigler et al.]
{\parbox{\textwidth}{\centering M.Reinbigler$^{1,2}$, R.Rouffet$^{2}$, P.Naylor$^{3}$, M.Czerkawski$^{3,4}$, N.Dionelis$^{3}$, E.Brunet$^{1}$, C.Fetita$^{5}$ and R.Martin$^{2}$  
        }
        \\
{\parbox{\textwidth}{\centering $^1$SAMOVAR, Inria Saclay, Télécom SudParis, Institut Polytechnique de Paris, Palaiseau, France\\
         $^2$Adobe Research, Paris, France\\
        $^3$  ESA/ESRIN, $\phi$-lab, Frascati, Italy\\
        $^4$  Asterisk Labs, London, UK\\
        $^5$ SAMOVAR, Télécom SudParis, Institut Polytechnique de Paris, Palaiseau, France
       }
}
}

\begin{document}

\definecolor{MarieColor}{RGB}{170,46,200}
\definecolor{RosalieColor}{RGB}{0,140,50}

\definecolor{RevOneColor}{RGB}{170,46,200}
\definecolor{RevTwoColor}{RGB}{0,140,50}
\definecolor{RevThreeColor}{RGB}{180,140,0}
\definecolor{RevFourColor}{RGB}{0,140,180}
\definecolor{RevFiveColor}{RGB}{180,0,150}

\colorlet{PURPLE}{purple}
\colorlet{BLACK}{black}

\newcommand{\warn}{\color{red}}
\newcommand{\warnoff}{\color{black}}

\newcommand{\revisionon}{\color{black}}
\newcommand{\revisionoff}{\color{black}}

\newcommand{\marieon}{\color{black}}
\newcommand{\marieoff}{\color{black}}
\newcommand{\marie}[1]{\textcolor{MarieColor}{M: #1}} 

\newcommand{\rosalieon}{\color{black}}
\newcommand{\rosalieoff}{\color{black}}
\newcommand{\rosalie}[1]{\textcolor{RosalieColor}{R: #1}} 

\newcommand{\ie}{\emph{i.e.}}
\newcommand{\eg}{\emph{e.g.}}
\newcommand{\etal}{\emph{et al.}}

\newcommand{\revOne}{\textcolor{RevOneColor}{1409}}
\newcommand{\revTwo}{\textcolor{RevTwoColor}{2312}}
\newcommand{\revThree}{\textcolor{RevThreeColor}{5495}}
\newcommand{\revFour}{\textcolor{RevFourColor}{0219}}
\newcommand{\revFive}{\textcolor{RevFiveColor}{1453}}

\teaser{
 \includegraphics[width=0.85\linewidth]{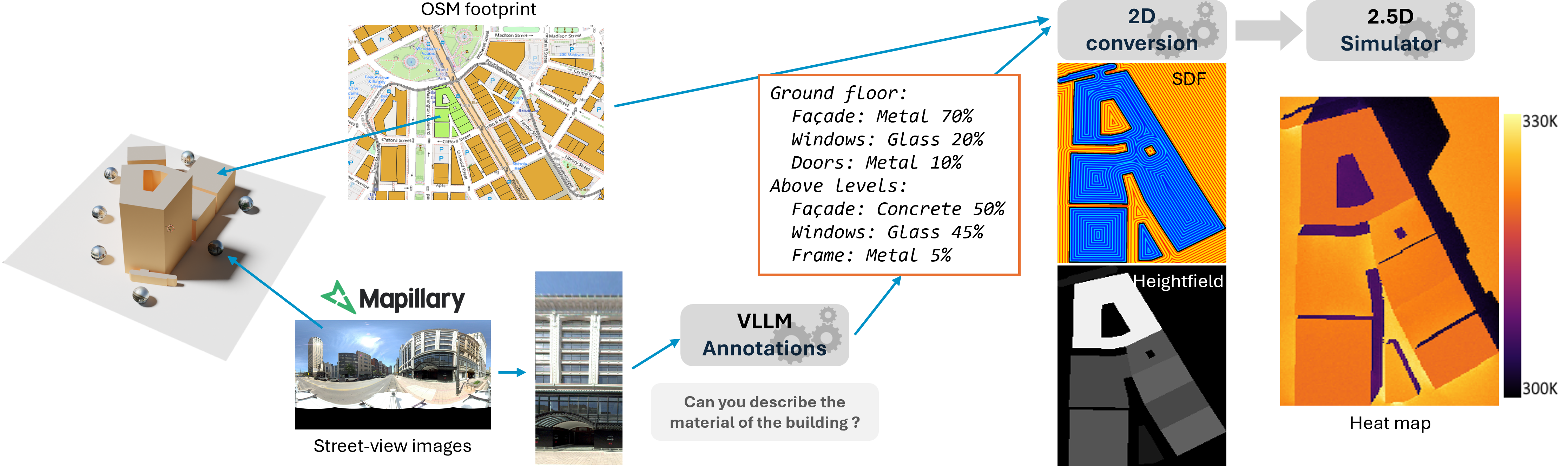}
 \centering
 \caption{We propose a pipeline for estimating the facade composition of buildings in a city, linking this information with their thermal properties to feed our 2.5D heat transfer simulator. First, we combine OpenStreetMap data with geolocated street-view 360° images to identify the best image for each building, which we input into a VLM to query the facade composition. We represent the city fully, including its materials and geometry, in randomly accessible 2D maps, which serve as inputs for our simulator that models coupled heat transfers.}
 \label{fig:teaser}
}

\maketitle

\begin{abstract}
The Urban Heat Island (UHI) effect, defined as a significant increase in temperature in urban environments compared to surrounding areas, is difficult to study in real cities using sensor data (satellites or in-situ stations) due to their coarse spatial and temporal resolution. Among the factors contributing to this effect are the properties of urban materials, which differ from those in rural areas. To analyze their individual impact and to test new material configurations, a high-resolution simulation at the city scale is required. Estimating the current materials used in a city, including those on building facades, is also challenging. We propose HeatMat, an approach to analyze at high resolution the individual impact of urban materials on the UHI effect in a real city, relying only on open data. We estimate building materials using street-view images and a pre-trained vision-language model (VLM) to supplement existing OpenStreetMap data, which describes the 2D geometry and features of buildings. We further encode this information into a set of 2D maps that represent the city's vertical structure and material characteristics. These maps serve as inputs for our 2.5D simulator, which models coupled heat transfers and enables random-access surface temperature estimation at multiple resolutions, reaching an x20 speedup compared to an equivalent simulation in 3D.


\end{abstract}

\section{Introduction}
\label{sec:intro}

The Urban Heat Island (UHI) effect designates the temperature gap between an urban area and its surrounding rural area. As a consequence of global warming, the UHI is amplified, both temporarily and in intensity. For example, by considering the 65 largest cities in the USA in 2024, 33.8 million Americans were exposed to UHIs with a gap in temperature of 8°F or more \cite{USA-UHI:2004}. Overheating impacts both the environment by disturbing and degrading it, and human health by increasing mortality and the occurrence of respiratory diseases. Studying this phenomenon and its causes is therefore essential.
Among the causes, urban materials' thermal properties are known to play a significant role in heat retention, which are adjustable parameters. \revisionon{}Thus, we focus this study on providing city planners with insights into their material selection independently from the other causes\revisionoff{}.

To study the UHI effect, surface temperature can be monitored thanks to remote sensing measurements covering the entire Earth, but at a low spatial resolution (100m/pixel) and very low temporal resolution (one data point every few days). The air temperature can also be an indicator, but is captured by very scarce meteorological stations. If it is sufficient for capturing and predicting the global heat flux, it is not accurate enough to study how urban materials locally affect the surface temperature and reinforce the UHI effect.

High-resolution surface temperature prediction can be performed using heat transfer simulators that evaluate the impact of different variables of the system in the resulting heat maps. Software based on physical equations solvers, such as ENVI-met \cite{ENVIMet}, has proven efficiency in simulating heat transfers in the field of city digital twins, as well as Monte Carlo based approaches, such as Stardis \cite{Stardis:2024}, but they both present one main drawback: the need for a 3D digital-twin model with material descriptions, which is a tedious task to obtain at the scale of an entire city. 
On the other hand, OpenStreetMap (OSM) \cite{OSM:2004} database contains the vector graphics description of the 2D maps of worldwide cities with a large coverage and easy to query via an API. Street-view images are omnipresent in dense cities, and can be used to estimate the facade materials per building, for a precise and automated way to describe the urban materials. 

To remove the need of a 3D digital twin, we propose to encode the city in \revisionon{} a set of top-view\revisionoff{} 2D maps \revisionon{}including the description of its verticality and material information to form a 2.5D city model. It is\revisionoff{} coupled with a novel, Monte Carlo based, heat transfer simulator that runs as a pixel shader on a GPU, leveraging our 2.5D city representation. Our 2.5D simulator follows the line of work of \cite{Caliot:2024} and \cite{Bati:2023} for coupled heat transfers, with direct and indirect solar irradiative, radiative, convective, and conductive transfers, with a 20x speedup and comparable results.


Our contributions can be summarized as follows: 

\rosalieon{}
\begin{itemize}  
    \item \textbf{HeatMat}: A novel approach allowing the study of the impact of materials on the UHI effect in a real city using only 2D data (OSM and street view images).
    \item A 2.5D heat transfer simulator running on GPU implementing coupled heat transfers (direct and indirect solar irradiative, radiative, convective, and conductive transfers) based on a Monte Carlo method that produces top-view heat maps of the surface temperatures, given the 2.5D city representation \revisionon{} and its location on Earth, the date, time and ambient temperature. We achieve a 20x speedup with comparable results to SOTA.\revisionoff{} 
    \item A 2.5D representation of a city, accounting for its geometry (height, signed distance field), and materials (top view, and facade materials procedural representation), via a set of textures, randomly-accessible from a GPU shader.
    \item A VLM-based pipeline for estimating the facade materials statistics of buildings from street views, enabling a procedural representation of the facade as a complement of OSM building data. 
    \item Validations of the heat transfer implementations involving conductive transfer, individually or coupled, with two different implementations, namely the Finite Difference Method and the Monte Carlo based Stardis prototype, as well as with satellite measurements.
    
\end{itemize}

\rosalieoff{}

\section{Background and Related works}  
\label{background}


    \subsection{Heat transfer simulation}
    
        The heat transfers can be simulated using two families of methods.

        \textbf{Deterministic methods.}
            Numerical methods based on physical equation solving have been extensively used to simulate heat transfer in urban areas. These are based on the discretization of the 3D environment. Tools are available at different scales: Urban canopy models \cite{UrbanCanopySnow:2016, TEB:2000, MultiLayerUrbanCanopy:2001, UrbanMesoscaleModel:2002, ENVIMet} \revisionon{}\cite{Umet:2008, WRF:2019, Urbanflow:2023} \revisionoff{}, and Building energy models \cite{BEM_Nouidui:2014, BEM-Beausoleil:2014, BEM-Crawley:2008, BEM-esru:2003, BEM-Katal:2019, BEM-Luziani:2019, BEM-Mahmoud:2020} \cite{BEM-Peng:2014, BEM-Roberts:2001, BEM-Robinson:2009}. However, these require meshing, and a predefined level of detail to build a model: e.g., a whole coarse city model, \ie{}, urban canopy model, or a detailed building model. The combination of model scales leads to approximations.



            \textbf{Monte Carlo methods} 

            \marieon{}       Using Monte Carlo methods for heat transfer simulation has been studied to model radiative transfers, leveraging photon tracing techniques \cite{Freude:2023, Freude:2025}. It can be combined with a differentiable modelization to optimize certain parameters, such as the new building geometry choices presented in \cite{Freude:2025}. On the other hand, individual conductive transfer equations solving has been implemented and recently improved by the introduction of the Walk on Stars \cite{Sawhney:2023, Miller:2024} that ensures only one boundary intersection per walk step, thus improving results quality. It is compatible with Dirichlet and Neumann boundary conditions \cite{Sawhney:2023} and further extended to Robin boundary conditions \cite{Miller:2024}. 
            \cite{Tregan:2023} has demonstrated theoretically the possibility of combining individual transfer models in one random path to couple radiative-convective-conductive transfers, which is also compatible with computer graphics techniques for efficient transfer solving. In addition, \cite{ibarrart:2022} proved the independence of these methods from the geometry size.             \revisionon{}These properties enable complex physical phenomena modeling, such as combining an atmospheric model with an urban model, while preserving complex multiscale scenes \cite{Teapot:2022}.
            \revisionoff{}\cite{Stardis:2024} implemented the theoretical work of \cite{Tregan:2023}. \cite{Bati:2023} added a post-mortem replay of the simulation, and \cite{Caliot:2024} added the solar irradiation contribution. 

            \marieoff{}

    \marieon{}             Our work proposes to leverage the Monte Carlo based model from \cite{Bati:2023} and \cite{Tregan:2023} to build a coupled heat transfers simulator, but instead of relying on a 3D city representations, it inputs 2D random-accessible maps describing the city, implying adaptations of conductive transfers from 3D to 2D.
    \marieoff{}

                   
    \subsection{Material retrieval}             
            Fine-grained material estimation that recovers the Spatially Varying Bidirectional Reflectance Distribution Function (SVBRDF) of a quasi-planar surface has been explored, relying on neural architecture to solve this ill-posed problem. In the literature, researchers have imposed lighting constraints with flash-light \cite{DESCHAINTRE1, DESCHAINTRE2, LuoX2024, JiaminCheng2024} and used in the wild image capture \cite{MaterIA:2022, GiuseppeVecchio, GiuseppeVecchio2}, or both \cite{XilongZhou, Sartor:2023, Zhu:2025}. 
                   
            For large-scale material estimation, satellite, aerial, and street view images are often used. \cite{ijgi7090339} generates a 3D city model that is automatically textured, also performing automatic building reconstruction. However, it requires the acquisition of airborne images at a $10$ cm resolution or more and a flight altitude of $800$ to $2500$ m. Recent papers detecting materials use several sources, including LIDAR \cite{LidarMaterialDetection:2022}, satellite for roof material detection \cite{RoofMatSatellite:2021}, and manual surveys. Automated methods at large scale exist and are based on ground views: (i) Google photorealistic tiles, proving only RGB textures \cite{GoogleTiles}; (ii) LuxCarta geospatial training and simulation products \cite{LuxCarta}; (iii) Geometry from Street-View \cite{StreetView3D:2016}; and (iv) \cite{BuildingMatDetection:2023} using street-view images and Computer Vision. However, \cite{BuildingMatDetection:2023} does not statistically represent facade material distribution, and thus requires training to extend to new cities. In this paper, we focus on the physical properties of materials and on a statistical facade representation that allows procedural element arrangement and connection to appearance and/or a physical material database.

    \subsection{Procedural facade}            

        Due to the inherent regularity and repetition of building facades, procedural modeling has been an effective paradigm for representing architectural facades. 
        Split grammars and L-systems have been extensively explored to represent city elements \cite{GrammarCity:2001,Wonka:2003,Muller:2006,GraphGrammar:2019}, further extended with inverse procedural modeling ability \cite{Muller:2007,Stava:2010,Wu:2014,SurveyModelingVirtualScenes:2024}. 
        More recently, the emphasis has shifted towards hybrid neuro-symbolic models that combine the structural interpretability of grammars with the expressivity of data-driven learning \cite{Plocharski:2024,plocharski:2025}. 
        
        Our work hierarchically combines simple shape elements to model the facade procedurally, and could be compatible with more advanced approaches.


    \subsection{City twin for UHI detection}                                                         
    Remote sensing data are extensively used to model the UHI effect. \cite{MaterialEffect-Small:2006} uses Landsat data to model the relationship between surface reflectance (how surfaces interact with light) and thermal conditions (surface temperature) across urban environments. Recently, remote sensing data are more commonly used, such as using LiDAR data to approximate 3D urban morphology~\cite{NewMat-Asadi:2020,Bagyaraj:2023}, Sentinel-2 data for Normalized Difference Vegetation Index~\cite{NewMat-Asadi:2020,Bagyaraj:2023}, or Landsat for approximating land surface temperature~\cite{NewMat-Asadi:2020,Bagyaraj:2023, land12061154, KAMARAJ2021112591}. Microclimate Vision~\cite{microclimate-vision}, integrates different sources of data, including satellite observations, to predict high-resolution microclimate data for cities~\cite{microclimate-vision}.
        
    UHI detection using city twin is exploited by \cite{ENVIMet}, which \revisionon{models many more physical phenomena than heat transfers. It is based on a voxel representation.\revisionoff{} It can also import OSM geometries to be extruded and voxelized but with aliasing effects. The ENVI-met software has been adopted by studies to analyze urban design and geometry influence on urban heat~\cite{NewMat-Ridha:2018, MaterialEffect-Acharya:2021,MaterialEffect-Tabatabaei:2023}, \revisionon{}but its radiosity model involves slow execution and in a very restricted area in the free version\revisionoff{}. Computational fluid dynamics based models are also proposed to analyze albedo influence on urban heat~\cite{MaterialEffect-Lopez:2022}. Some models simulate heat transfers coupled with atmosphere effects, e.g., (i) \cite{NumMethCoupledTransfers:2023} includes a mass transfer model in addition to coupled heat transfers to include a climate model, and (ii) \cite{TEB:2000} 
    models energy transfers in a simplified 3D city geometry, including an atmospheric model.
   

    \revisionon{}In this work, we propose to isolate the effect of materials on the UHI phenomenon\revisionoff{} by performing a high-resolution simulation of the heat transfers from the city materials and geometry description extracted automatically from open-source data.
\section{City acquisition}
\label{City acquisition}
\begin{figure*}[t]
    \centering
    \includegraphics[width=1\textwidth]{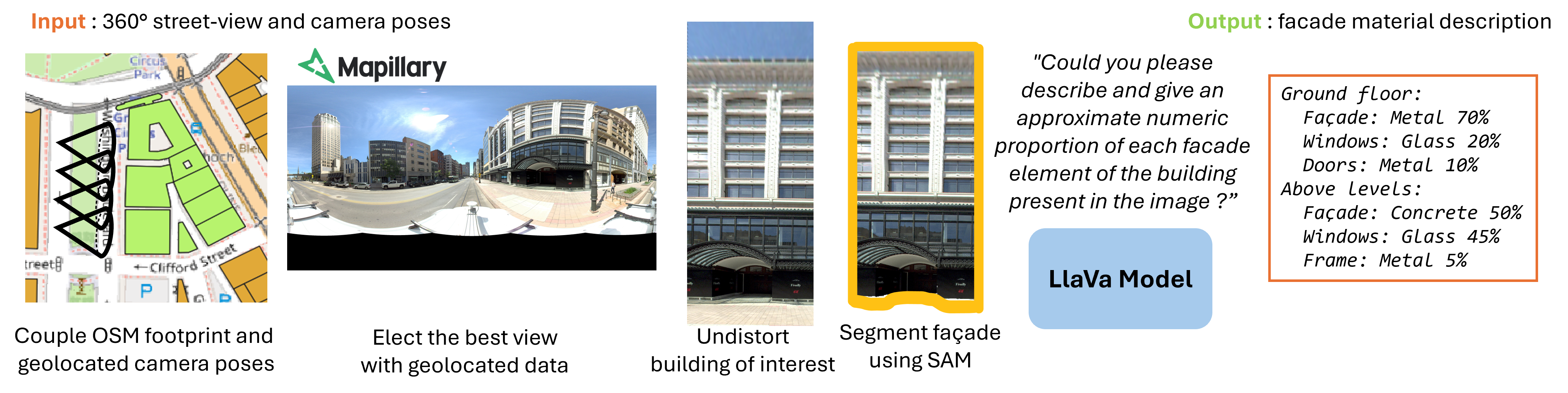}
    \caption{Pipeline for material retrieval. From left to right: Starting with street-view 360° images and geolocated camera positions, we couple both information and elect, for each building, the 360° image that best represents it. We rectify this image and extract the building facade of interest, segment it, and query a VLM to estimate the material and proportions of each element in the facade.}
    \label{fig:VLM annotation}
\end{figure*}
As highlighted by previous works, Monte Carlo methods are commonly adopted to solve heat transfers in an urban geometry, leveraging rendering techniques thanks to their ability to scale to the geometry size. However, current implementations require a 3D representation of a city, which is a tedious task at the scale of a real city. We propose to rely on 2D information only, OpenStreetMap data for the city map, coupled with street-view images (Sec. \ref{street-view OSM coupling}), from which we estimate the facade materials using a VLM-based pipeline (Sec. \ref{VLM-based pipeline}). We encode the city into \revisionon{}18 2D textures\revisionoff{} (Sec. \ref{Top_view_maps}), including the facade \revisionon{}description\revisionoff{} (Sec. \ref{Facade_maps}), to allow for running the simulation as a simple pixel shader and gain performance. \revisionon{}The whole pipeline is summarized in Fig. \ref{fig:VLM annotation}\revisionoff{}.



\subsection{Coupling street-views with building footprints}
\label{street-view OSM coupling}

Several open data sources are available, including Mapillary, which provides both user-captured street-view images as well as a set of curated datasets. Among these, we select the Metropolis dataset \cite{Metropolis:2024}, captured in Detroit, Michigan, with a 360° camera on a geolocated car. As we focus on buildings, we use the geolocation metadata to deduce the field of view (FOV) cone of each subsection corresponding to the left and right views from the car position, likely to contain buildings. We query the OSM database with this cone, with a far distance large enough to cross a standard street, to retrieve all building footprints intersecting it. We discard from the query results the buildings fully occluded by the closest buildings to the camera position. 
We then augment the Metropolis metadata with, for each captured view, the visible buildings, their rotation, as well as their relative position, and coverage in the image. We further use this information to identify the best view of each building, by scoring the building's proximity to the camera, frontal positioning and coverage. 
This allows for a reduction by a factor of 10 in the number of images to process, for a sober use of the following material estimation pipeline. 

\subsection{VLM-based facade composition estimation}
\label{VLM-based pipeline}

We propose a pipeline for estimating the materials in a facade using image processing and state-of-the-art pre-trained models. It starts with the recovery of the left and right rectified pictures from the 360° equirectangular views from Mapillary, using the Perspective Fields method \cite{jin2022PerspectiveFields}, computed for the four azimuth angles (0,~90,~180,~270) of the panorama and averaged to uniformly rectify it, accounting for an 85° vertical field of view. Only the left and right images are kept from the rectification, and each image is then sliced to extract only the building view corresponding to our metadata. Each sliced image is then segmented using GroundedSAM \cite{GroundedSAM:2024} that we query with the prompt "facade" to keep only the buildings. We crop the segmented image to the bounding box of the mask, to focus the VLM attention. 

The segmented image is provided as input to a pretrained VLM model, Llava 1.6~\cite{Llava1.5:2023} in our experiments, which can be easily replaced. We design our prompt so that the VLM returns, in JSON format, the material and proportion of each important facade element, separated for the ground level (main material, doors, windows, frames) and the upper levels (main material, windows, frames, shutters). We augment our building metadata with this data.

For validating the facade materials and proportions estimated, we manually built a ground truth dataset made of 50 street views that represent one building facade, for which we label the materials and proportions manually, by carefully analyzing the image. An ablation of the use of GroundedSAM, and measures of the material estimation with IoU and accuracy metrics, using three different VLMs, are presented in the Suppl. Table \ref{table_vlm_validation}.


\subsection{Top view data encoding into texture}
\label{Top_view_maps}

The first set of maps describes the features of the building visible from a top view, to be stored in a 2D map of the city. Among those features, there are the identifier (id) of each building, its height, and the material composition of the roofs and the ground. 
We create the maps by applying a set of operations on the footprint polygons using the building metadata. We rasterize the area of interest with a cell size of the resolution of the simulation chosen in meters. When a cell is at the boundary of multiple buildings, we store the information of the building with the largest overlap. We pack our scalar data into RGBA 16-bit textures to minimize the texture sampling later in the simulator.



\subsection{Facade representation and encoding}
\label{Facade_maps}


To represent the verticality of the city, not visible from a top view, we define a methodology to first encode the facade geometry while removing the aliasing effect due to the rasterization of the footprint polygons for a more accurate intersection detection. 

Thus, we represent the facade orientation using the couple (distance, normal). The distance corresponds to the space between the pixel centroid and the closest facade, which forms a discrete SDF to be used for Sphere Tracing. The normal corresponds to the facade normal, expressed as the azimuth angle of the facade. To ensure one normal and one distance per pixel for the external facade, we modify the geometry by approximating corners by one segment connecting the two points of intersection to the grid cell. Figure \ref{fig:aliasing solution} illustrates this approximation to preserve intersection continuities.

\begin{figure}   
  \centering 
    \includegraphics[width=0.5\textwidth]{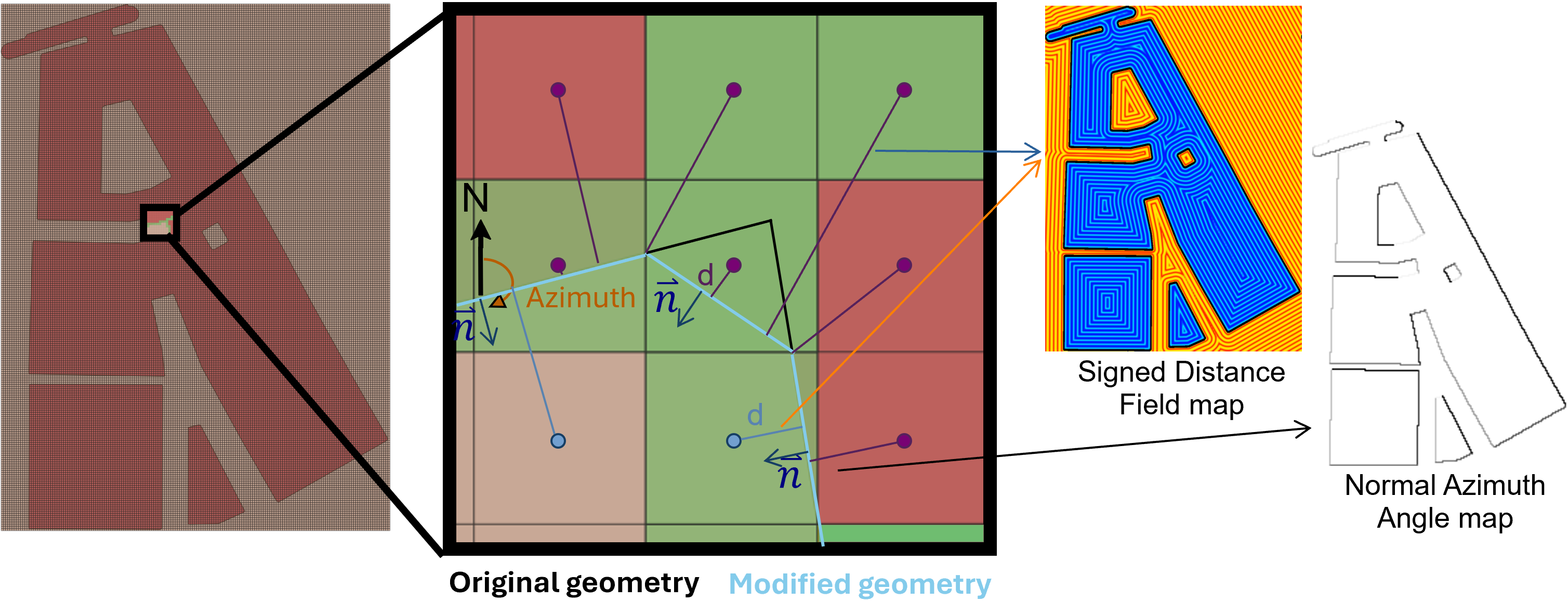}
    \caption{Illustration of the rasterization of footprint buildings. In each pixel rasterized (left) we keep only one segment that represents a facade, to allow for the discrete SDF representation (right). At the corners (middle) we simplify the geometry to connect the two segments coming from the adjacent facades.}
    \label{fig:aliasing solution}
\end{figure}

To ensure coherence in the sampled facade material, we build a procedural function that associates with each position (u,h) in the facade plane the building component to which it belongs (door, window, etc.), further matched with the corresponding material in that facade (metal, glass, etc.). To allow a random accessible way to get the relative position u along a facade plane, we parameterize it as an interpolation between 0 and 1, during the polygons rasterization. At simulation time, when querying the facade material at a point P(x,y,z), h is given by P.z/height, and u is obtained by adding to the queried u value of the u map the facade length, divided by the whole perimeter length. Details are available in Fig.~\ref{fig:umap_explanation}. Using a procedural sampling function, based on the 2.5D encoded material maps, we identify the facade component to which our point belongs and the material \revisionon{}id\revisionoff{}. 

\begin{figure}   
  \centering 
    \includegraphics[width=0.48\textwidth]{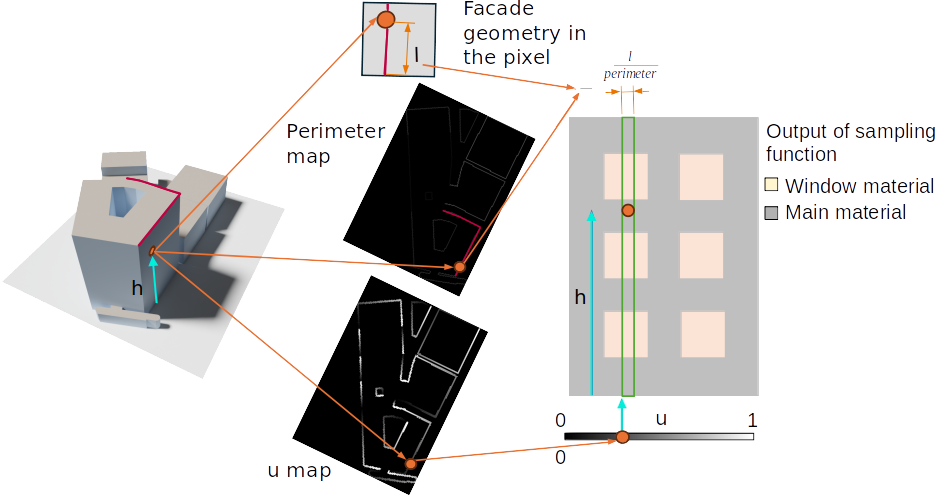}
    \caption{The u map and perimeter maps allow for retrieving the relative position of a 3D point in the plane of a facade, and for guaranteeing a consistent facade representation in the simulator.}
    \label{fig:umap_explanation}
\end{figure}

\revisionon{}To retrieve the facade material properties\revisionoff{}, we use the 8 first bits for the material id and the 8 following bits for the percentage. Table 1 in Supplemental Material presents the material thermal properties database, aggregated from several sources \cite{MaterialLitt1, MaterialLitt2, MaterialLitt3, MaterialLitt4, MaterialLitt5, MaterialLitt6, MaterialLitt7}.

2.5D city maps and a visual of the procedural facade sampling are illustrated in Fig.~\ref{fig:procedural_facade}.
\revisionon{}This facade sampling, coupled with the procedural representation of facade and the 2D random-accessible geometry encoding (Umap, perimeter map, height map, SDF, and material information) allows to simulate different facade elements along the vertical axis, which is novel to the best of our knowledge.\revisionoff{}

\begin{figure}[ht]   
  \centering 
    \includegraphics[width=0.5\textwidth]{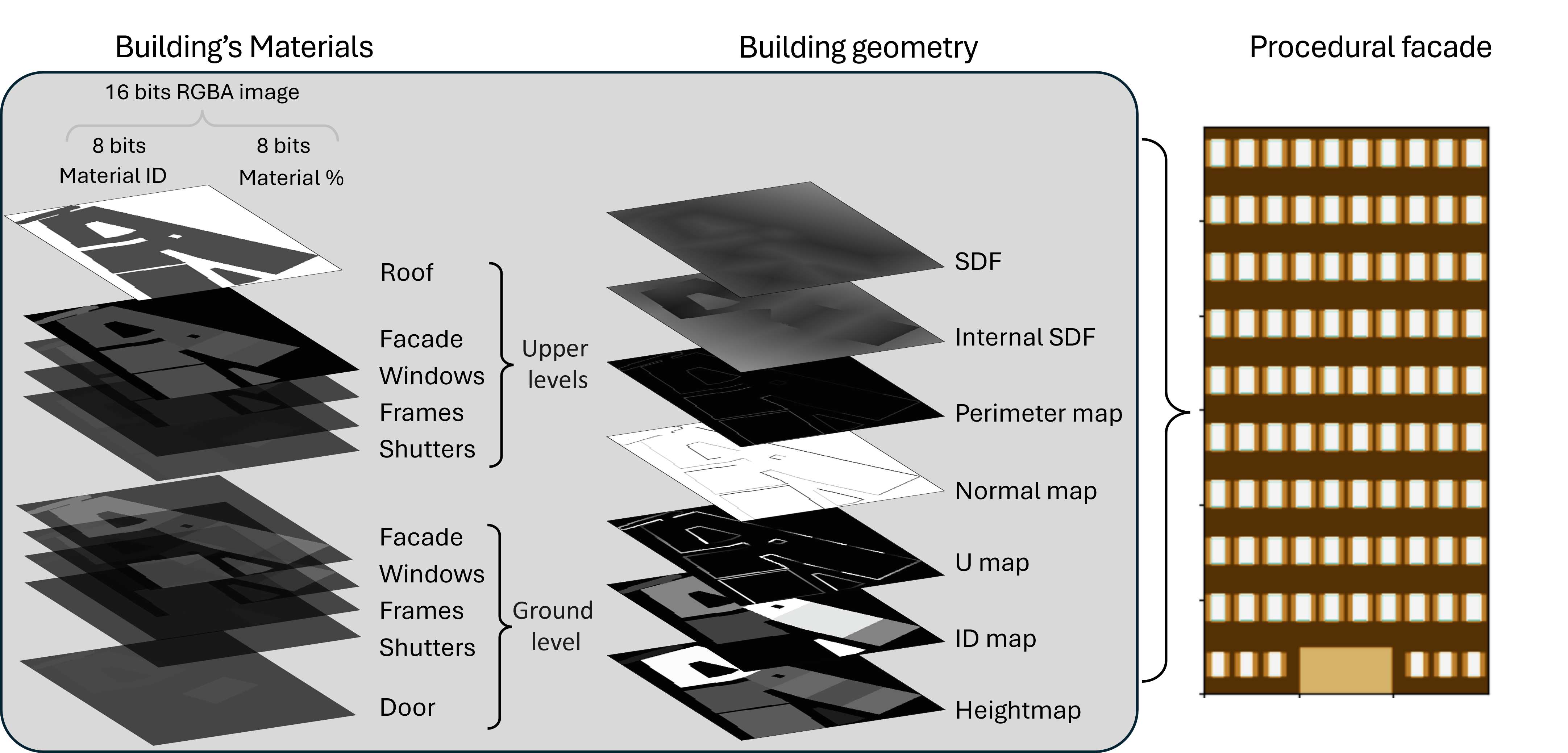}
    \caption{2.5 representation of a city using a set of maps, with a top view (roofs, ground), building facade materials (left), city geometry height map, signed distance field, and facade normals (middle). It enables procedural modeling of a facade (right) when intersecting points on a facade in the heat transfer simulator. }
    \label{fig:procedural_facade}
\end{figure}

\section{2.5D heat transfer simulation}
\label{Simulation description}


The set of 2D maps generated making this information random-accessible, offers the opportunity to build a simulation running on a GPU as a pixel shader. Thus, it takes advantage of computer graphics techniques, such as sphere tracing and walk-on-sphere, enabling an efficient intersection computation. The simulation, based on Monte Carlo methods, computes the heat transfers in an urban area, including direct and indirect solar irradiation, radiative, convective, and conductive heat transfers. Each one is solved independently as described in Sec.\ref{Individual transfers design}. The \revisionon{}main\revisionoff{} mathematical and physical equations \revisionon{}presented hereafter are extracted from\revisionoff{} Caliot \etal{}'s paper\cite{Caliot:2024}, including how Monte Carlo methods and computer graphics techniques can solve the problem in a 3D geometry. 
Our goal is to estimate the surface temperature $T_s$ using Monte Carlo estimator $\widetilde{T}_s(\vec{y}_0, t_0)$ where 
\vspace{-5 pt}
\begin{equation} 
T_s(\vec{y}_0, t_0) \approx \widetilde{T}_s(\vec{y}_0, t_0) = \frac{1}{N} \sum_{k=1}^{N} W_k
\end{equation}
$W_k$ is the Monte Carlo method weight of one realization defined in Eq.~\ref{eq:MCweight}

This weight can be decomposed as the sum of the possible path end temperatures, to \revisionon{}which the weight $W_{o,k}$ is added\revisionoff{} (Eq.~\ref{eq:MCweight}). The weight contains $T_I$ when the path reaches the initial time \revisionon{}$\tau_I$\revisionoff{} via the conductive transfer. Otherwise, it is composed of $T_D$ when it reaches a Dirichlet boundary, a known solid temperature, $T_F$ at a solid-fluid interface corresponding to the fluid temperature involved in the convective transfer or $T_{sky}$ when a radiative ray ends in the sky. \revisionon{}$\delta_{condition}$ designates the Dirac function : it is worth 1 if the condition is met, 0 otherwise.\revisionoff{}

\vspace{-9 pt}
\begin{align}
\label{eq:MCweight}
    W_k(\vec{y}_0,t_0) &= \delta_{t_n \leq \tau_I} T_I + \delta_{t_n > \tau_I} \nonumber \\ & \Bigg\{ 
        \delta_{\vec{y}_n \in \partial D_S} T_D(\vec{y}_n, t_n) \nonumber \\
        & + \delta_{\vec{y}_n \in \partial D_{S,F}} T_F(\vec{y}_n, t_n) \nonumber \\
        & + \delta_{\vec{y}_n \in \partial D_{sky}} T_{sky}(\vec{y}_n, \vec{\omega}_n, t_n) \Bigg\} \nonumber \\ &+ W_{o,k}
\end{align}

$W_{o,k}$ refers to the solar contribution to the temperature estimation. It is composed of two parts: the direct solar irradiation $W_{o,d}$ and a second one corresponding to the indirect contribution along the indirect path (Eq.~\ref{eq:Wok_def}). \revisionon{} $n_{o}$ designates the number of Robin boundary conditions met, $n_{r,j}$ to the number of reflections on the indirect path and $h_T$ the sum of each heat transfer mode coefficient.\revisionoff{}

\vspace{-9 pt}
\begin{align}
    \label{eq:Wok_def}
W_{o,k} &= \frac{1}{h_T} \sum_{j=1}^{n_o} 
\Bigg\{ W_{o,d}(\vec{y}_j) \nonumber \\ &+
\left(\sum_{m=1}^{n_{r,j}} \delta_{\vec{y}_m \in \partial D_{S,F}^{L}} W_{o,r,d}^{L}(\vec{y}_m) \right) \nonumber \\ &+
W_{o,sky}(\vec{y}_{n_{r,j}+1}) + W_{o,r,d}^{F}(\vec{y}_{n_{r,j}+1}) \Bigg\}.
\end{align}

The direct solar radiation is defined in Eq.~\ref{directsunMcweight} as a function of the solar direction and normal vectors $\vec{\omega}_d$ and $\vec{n}_0$, the direct solar irradiation $D_o$, and the material emissivity $\varepsilon$, \ie{} material's ability to radiate absorbed heat. Based on Kirchhoff's law, $emissivity = absorptivity$ for all wavelengths, \ie{} for opaque material $emissivity = 1 - albedo$.

The indirect solar irradiation contribution can be decomposed into three components: the diffusion of solar irradiation at each bounce on a Lambertian surface (Eq.~\ref{lambertianWeight}), and two possible path end conditions. One end condition corresponds to an indirect path ray reaching the sky, thus adding a weight defined by Eq.~\ref{indirectSkyWeight}, modeling an atmospheric sky model where it is proportional to the direct radiative solar intensity $I_{o,d}$. The second end condition corresponds to the reflection of the ray on a specular surface with a reflected ray contained in the sun half-angle. It depends on the material emissivity and the direct radiative solar intensity  $I_{o,d}$ (Eq. \ref{specularWeight}).

\vspace{-15pt}
\begin{align}
    \label{directsunMcweight}
    & W_{o,d}(\vec{y}_0) = \delta_{\vec{y}_d \in \partial D_d} \varepsilon(\vec{y}_0) |\vec{\omega}_d \cdot \vec{n}_0| D_o(\vec{y}_0) \\
    \label{lambertianWeight}
    & W_{o,r,d}^{L}(\vec{y}_i) = \varepsilon(\vec{y}_0) \delta_{\vec{y}_d \in \partial D_d} |\vec{\omega}_d \cdot \vec{n}_i| D_o(\vec{y}_i) \\
    \label{indirectSkyWeight}
    & W_{o,sky}(\vec{y}_i) = \varepsilon(\vec{y}_0) \pi \delta_{\vec{y}_{i+1} \in \partial D_{sky}} f_{sky}(-\vec{\omega}_i) I_{o,d}(\vec{y}_i) \\
    \label{specularWeight}
    & W_{o,r,d}^{F}(\vec{y}_i) = \varepsilon(\vec{y}_0) \pi \delta_{\vec{y}_i \in \partial D_{S,F}^{F}} \delta_{\vec{y}_{i+1} \in \partial D_d} I_{o,d}(\vec{y}_i)
\end{align}

\revisionon{} Sections \ref{Individual transfers design} and \ref{Coupling} describe how we implemented the solving of these equations in practice in 2.5D, inspired by previous works \cite{Caliot:2024, Bati:2023} and including adaptations to the 2D conductive model and data representation specific to this paper.\revisionoff{}

\subsection{Individual heat transfers design}
\label{Individual transfers design}

\subsubsection{Conductive heat transfer}
\label{Conductive transfer}
 
In our 2.5D representation, \revisionon{}only top-view information are accessible which implies we do not model the building interiors\revisionoff{}. Thus, our conductive heat transfer implementation leverages the Walk-On-$\delta$-sphere, similarly to \cite{Bati:2023}, but in \revisionon{}2D dimension on surfaces, meaning across the facades, roofs, and ground surfaces\revisionoff{}. Like us, Caliot \etal{} and Bati \etal{} compute the conductive path based on the Walk-on-$delta$-Sphere techniques, \ie{} the Walk-on-Sphere method, but instead of estimating the radius of the sphere \revisionon{}when\revisionoff{} tangent to the surface, we move with a sphere with a fixed radius~$\delta$. Both followed Tregan \etal{} \cite{Tregan:2023} statement saying that the estimation matches accurately the exact solution when $\delta=\frac{L}{20}$, where L is the characteristic length of the problem. In the 3D context, the characteristic length of the conductive transfer is the wall thickness. To adapt it to our 2D case, \revisionon{}we propose to define the characteristic length as $min(width_{pattern}, height_{pattern})$ size where the pattern corresponds to the unit facade arrangement being used in the facade sampling\revisionoff{}. In addition, as we remove a dimension along the facade, the coupled conductive path algorithm requires adjusting the exponential sampling mean of the rewind in time as follows: $$ \frac{\rho c \delta^2}{6k} \longrightarrow \frac{\rho c \delta^2}{4k} $$
with $\rho$ the material density, c the heat capacity, and k the thermal conductivity.


\subsubsection{Radiative heat transfer and solar irradiation}
\label{Radiative transfer}

The radiative heat transfer is divided into two categories: the terrestrial radiative transfer generated by objects' temperature and the solar radiative transfer resulting from the sun irradiation. 
In both cases, we need to explore the space, which is implemented using the sphere tracing technique based on a discrete Signed Distance Field (SDF) stored in the SDF map used as input. The intersection computations are performed in 3D with a building geometry described by its heightfield and the facade description, as discussed in Sec.~\ref{Facade_maps}.
For the solar radiative transfer, to better represent reality, we \revisionon{}simulate the real sun path depending on time and date\revisionoff{}. 


\subsubsection{Convective transfer}
\label{Convective transfer}

The volume of air temperature in an urban area is substantial, the inertia required for the air temperature to be increased globally under the influence of the solids present in the scene is large. Thus, we can consider that the convective transfer contribution to the Monte Carlo estimator of the surface temperature equals the air temperature at the local time, similarly to Caliot \etal{} work \cite{Caliot:2024}.

\subsection{Heat transfers coupling}
\label{Coupling}

\begin{figure}[htb]
  \centering 
    \includegraphics[width=0.45\textwidth]{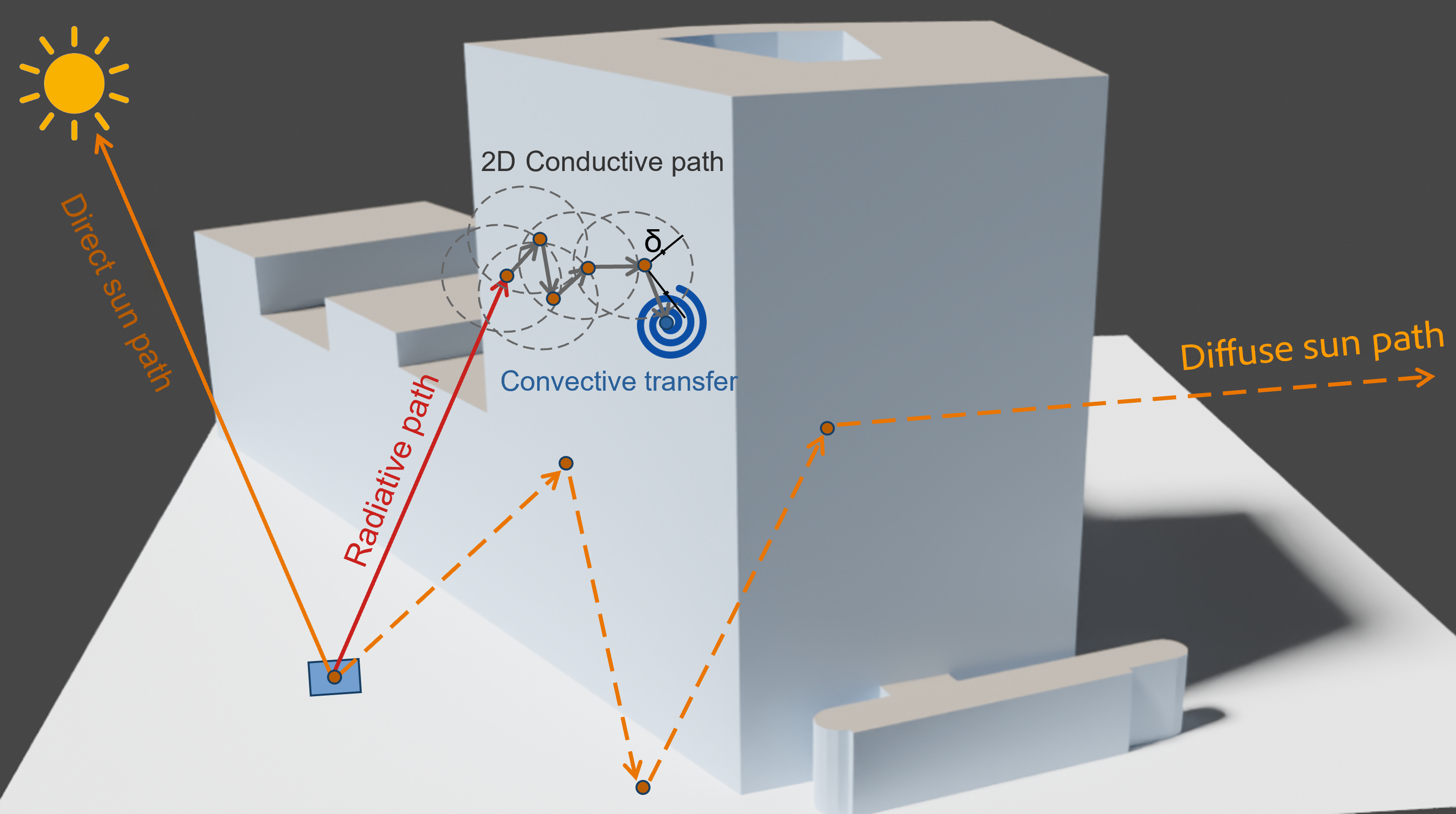}
    \caption{Coupling of radiative, convective, conductive, as well as solar direct and indirect irradiation.}
    \label{fig:coupledTransfer}
\end{figure}

The coupling, inspired by the Bati \etal{} pseudo-code \cite{Bati:2023}, is completed by the solar contribution described by Caliot \etal{}~\cite{Caliot:2024}, with adaptations due to our 2.5D setting. We start our computation at the location corresponding to each pixel centroid, at an altitude matching the height map value. Then, at each solid-fluid interface reached, we compute the solar irradiation contribution to the final temperature. When at a solid-fluid interface, we randomly select a transfer mode using the material properties dependent probability to select a radiative, a conductive, or a convective heat transfer as described in Caliot \etal{} \cite{Caliot:2024}. When a convective transfer is sampled, the path ends and returns the air temperature at the time reached by the simulation, plus the solar contribution accumulated along the path. If the radiative heat is sampled, then a ray is sent in a randomly sampled direction using a cosine distribution. If it hits a geometry, a new sampling of the heat transfer type is computed. Otherwise, the path is ended and returns the corresponding to the sky temperature, added to the accumulated solar Monte Carlo weight. Finally, if the conductive transfer is selected, we need to adapt it to our surface transfer instead of volumetric transfer. Consequently, as the conductive path evolves on the facade surface, it remains at the solid-fluid interface at each step. Thus, contrary to the literature method, we do not sample a new heat transfer type once we reach an interface, but after a number of steps to be parametrized. At each step of the conductive path, a rewind in time is performed accounting for the material property. If the initial time is reached, the conductive path ends the simulation path by returning the initial temperature at the current location together with the solar contribution. An illustration of all transfers at stake is available in Fig.~\ref{fig:coupledTransfer}.

\section{Evaluation and performances}
\label{evaluation}

We evaluate individually our 2D implementation of the conductive transfer compared to the Finite Difference Method (Sec.~\ref{Individual eval}), and we validate its coupling with radiative and convective transfer (Sec.~\ref{Coupling eval}) and against satellite measurements (Sec.~\ref{satellite_eval}). 

\subsection{Individual conductive transfer evaluation}
\label{Individual eval}


We compare our 2D implementation of the conductive transfer against the surface Finite Difference Method. The simulation consists of a Carbon 1\% Steel heated continuously at its two vertical extremities to reach 1°C and maintained at 0°C at the horizontal extremities. We compare the result at the stationary state. The initial temperature is set to 0°C in the whole rectangle except the top and bottom row, heated at 1°C. \revisionon{}Due to the method’s stochasticity, the results only approximate the solution, thus introducing small deviations. When computing the scaled difference as defined in \cite{Caliot:2024}, we observe an error of 1\% compared to the reference, which is in the same order of magnitude as reported in  \cite{Caliot:2024}.\revisionoff{}



\begin{figure}[htb]
  \centering 
  \vspace{-7pt}
  \begin{subfigure}[t]{0.21\textwidth}
    \includegraphics[width=0.80\textwidth]{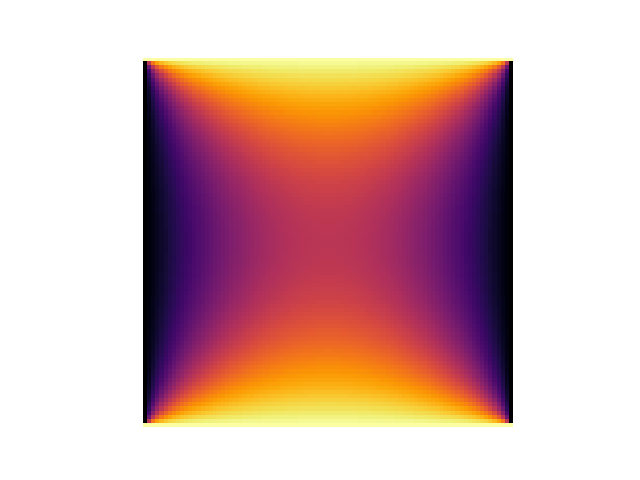}
    \caption{Finite Difference Method}
    \label{fig:FDM}
  \end{subfigure}
    \hfill
  \begin{subfigure}[t]{0.26\textwidth}
    \includegraphics[width=0.65\textwidth]{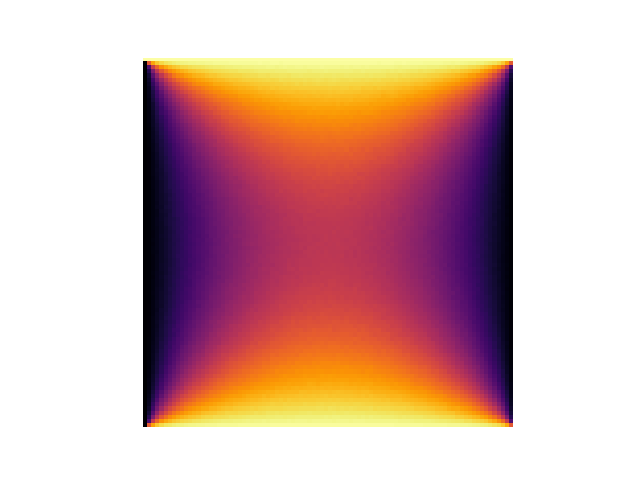}
    \caption{Our Walk-on-$\delta$-Sphere based method}
    \label{fig:Wos}
  \end{subfigure}
  \vspace{-7pt}
  \caption{Conductive transfer comparison at the stationary state presents a mean scaled difference of 1\%.} 
  
  \label{fig:conductive_eval}
   
\end{figure}


\subsection{Coupled heat transfers evaluation}
\label{Coupling eval}

\begin{figure}[htb]
  \centering 
  \begin{subfigure}[t]{0.235\textwidth}
    \includegraphics[width=\textwidth,height=3cm]{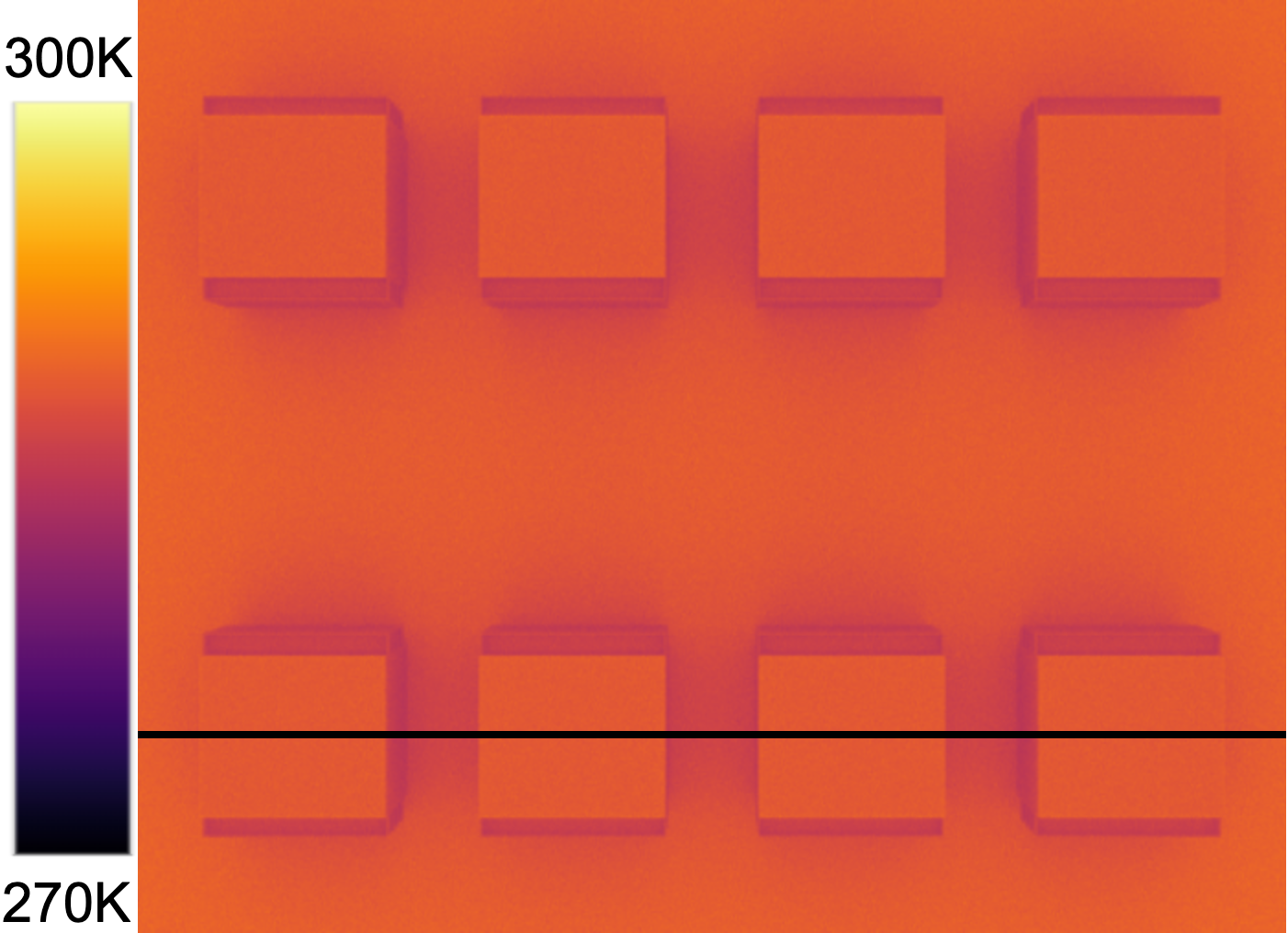}
    \caption{Stardis}
    \label{fig:stardis}
  \end{subfigure}
    \hfill
  \begin{subfigure}[t]{0.21\textwidth}
    \includegraphics[width=\textwidth,height=3cm]{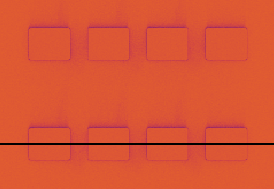}
    \caption{Our proposed method}
    \label{fig:our_stardis}
  \end{subfigure}
    \begin{subfigure}[t]{0.23\textwidth}
    \includegraphics[width=\textwidth]{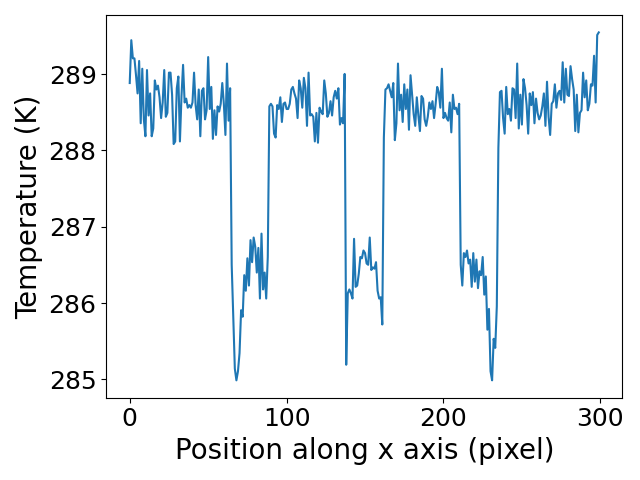}
    \caption{\revisionon{}Temperature profile of Stardis\revisionoff{}}
    \label{fig:stardis_profile}
  \end{subfigure}
    \begin{subfigure}[t]{0.23\textwidth}
    \includegraphics[width=\textwidth]{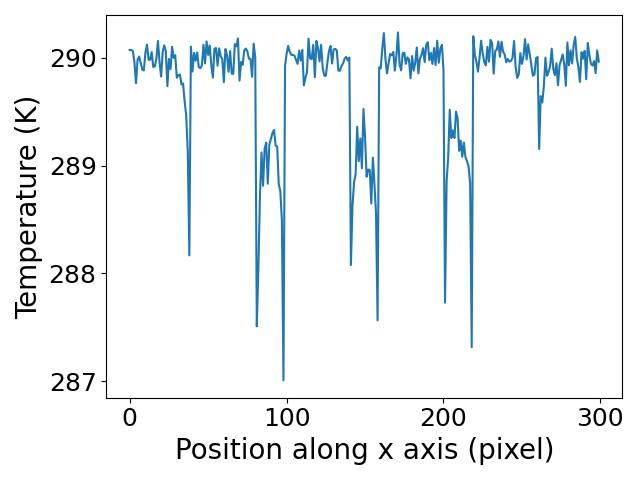}
    \caption{\revisionon{}Temperature profile of our output\revisionoff{}}
    \label{fig:our_stardis_profile}
  \end{subfigure}
  \caption{\revisionon{}Coupled heat transfer comparison, without solar irradiations, from a top view heatmap and via temperature profile along the displayed line.\revisionoff{} Our 2.5D method provides comparable results to Stardis, which runs on a 3D geometry.}  
  \label{fig:coupled_heat_transfers}
   \vspace{-7pt}
\end{figure}

We compare our coupling implementation to the available 3D Monte Carlo based simulator Stardis \cite{Stardis:2024}, which models the same transfers as our method except for the solar contribution. We set all initial conditions according to Suppl. Tab.~\ref{table_material_comp_stardis}. To be able to run the simulation on the same geometry, we created a 2.5D representation of \revisionon{}210x320 pixels of the Stardis starter pack with a 0.5m per pixel resolution. As for Stardis, the 3D model weighs 11.4MB while our corresponding 2.5D representation weighs 1MB.\revisionoff{} Figure~\ref{fig:coupled_heat_transfers} presents the top-view heatmap outputs in the same color range. Due to the rendered perspective and to balcony-related effect, as we do not model those elements, the resulting map cannot be superimposed on our heatmap. Thus, we estimate the average temperature per area of interest: roofs, ground far from buildings, and in-between buildings. The values deviate respectively from
0.1\%, 0.4\%, and 0.2\% from Stardis values, which correpond to less than a Kelvin, which should not affect the ability to detect heat islands perceptible by human beings.
\revisionon{}
The temperature profiles along the horizontal line shows the three down peaks, corresponding to the drop in temperatures between the building which are in the shadow. The two additional peaks in our result correspond to the left and right façades, hidden by the perspective on the Stardis result. The left drop is larger than the right one as it is less exposed to the sun. We can also note that our simulator tends to heat a bit more than Stardis but follows the same overall evolution.
\revisionoff{}

In Tab.~\ref{table_simulator_perfs}, we compare the performance of our 2.5D simulator against Stardis on this geometry, with varying samples per pixel (fewer samples, more noise) \revisionon{}on a laptop equipped with a Processor 11th Gen Intel(R) Core(TM) i7-11800H (2.30 GHz) and an NVidia GeForce RTX 3080 Laptop GPU\revisionoff{}. The 20x speedup is obtained by writing the simulator as a GPU shader, using random accessible maps for representing the city. \revisionon{}Our simulator scales linearly with the maps resolution and the number of samples per pixel.\revisionoff{} 


\begin{table}[hb]
\centering
\small
\vspace{-0pt}
\begin{tabular}{|lllll|}
\hline
Simulator / SPP & 125 & 250 & 500 & 1000 \\
\hline
Stardis     & 11min & 21min41  & 44min11 & 90min10 \\
Ours w/o irrad. & \textbf{42sec} &  \textbf{1min21} & \textbf{2min41} & \textbf{5min15} \\
\hline
\end{tabular}
\vspace{-6pt}

\caption{Simulator performances. Ours is 20x faster.}
\label{table_simulator_perfs}
\end{table}





\subsection{Comparison to Satellite Land Surface Temperature} 
\label{satellite_eval}

\setlength{\belowcaptionskip}{-1pt} 
\begin{figure}[htb]
  \centering 
  \begin{subfigure}[t]{0.2\textwidth}
    \includegraphics[height=2.3cm]{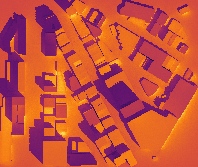}
    
    \caption{Detroit Heatmap (K)}
  \end{subfigure}
  \hfill
  \begin{subfigure}[t]{0.20\textwidth}
    \includegraphics[height=2.3cm]{figures/ComparaisonLandsat/DetroitDistrictGT_27072024_resized}
    \caption{Downsampled heatmap (K)}
    \end{subfigure}
  \hfill
   \begin{subfigure}[t]{0.2\textwidth}
    \includegraphics[height=2.3cm]{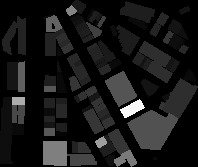}
    \caption{Detroit heightmap}
  \end{subfigure}
  \hfill
  \begin{subfigure}[t]{0.2\textwidth}
    \includegraphics[height=2.3cm]{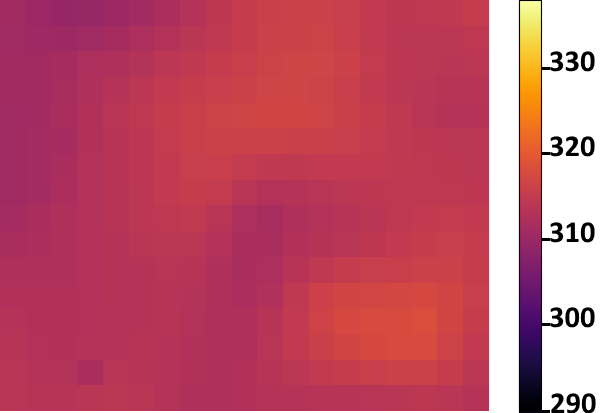}
    \caption{Landsat heatmap (K)}
  \end{subfigure}
  \hfill
  \caption{Detroit district simulation, July 27, 2024, at 12:15 pm with the same initial conditions as in Sec. 6. experiment. \revisionon{}The gradients of our downscaled simulation (b) are very similar to the Landsat capture (d) at the same day and time.\revisionoff{}}
  \label{fig:detroit_GT_comp}
\end{figure}

Collecting ground truth data of surface temperatures at high temporal and spatial resolutions is a complex task, which is why being able to simulate it is so crucial for providing urban planners with more insights. For instance, the Landsat satellite captures surface temperatures at 100m/pixel every 16 days. We found one record for Detroit around 12:16 pm on July 27, 2024, retrieved the ambient temperature at that time, and ran our simulator with these settings. We applied several concrete versions with different emissivities directly derived from the albedo estimation performed on satellite images. \revisionon{}We generate the set of 18 198x167 pixels maps with a spatial resolution of 3m/pixel for a broad overview.\revisionoff{}
Fig. \ref{fig:detroit_GT_comp} presents our simulation, and the Landsat record, \rosalieon as well as Fig.~\ref{fig:Detroit_Landsat_Comp} \rosalieoff.  Darkest roofs corresponds to lightest concrete with highest albedo, whereas hottest one corresponds to darkest concrete version. We performed bilinear downsampling and upsampling on our heatmap to enable a direct comparison at the same resolution. 
\marieon{}Considering that Landsat measurements are accurate at about 3K close and is lower in urban areas than in rural areas, our results in absolute value are in general slightly hotter but the overall gradient is very similar, confirming our intuition. 

Similar experiments have been performed on several other dates with low cloud coverage. Fig.~\ref{fig:Detroit_Landsat_Comp} illustrates that our simulator output follows similar heat distribution as the Landsat heatmap, 
The \revisionon{}shift in absolute values\revisionoff{} can be explained by the Landsat images themselves that do not include any radiometric calibration, atmospheric correction nor emissivity correction, which can introduce some \revisionon{}gaps\revisionoff{}. In addition, weather conditions, vegetation or anthropogenic activities are also not studied as we aim at isolating material impact independently of the other parameters. 
\marieoff{}

\begin{figure}[htbp]
\centering
\setlength{\tabcolsep}{0.1pt}     
\renewcommand{\arraystretch}{1} 
\begin{tabular}{ccccc}
\hfill
\textbf{Mar 13} & \textbf{Apr 14} & \textbf{Jul 27} & \textbf{Aug 12} & \textbf{Sep 21} \\
\hfill
19\textdegree{} & 25\textdegree{} & 26.5\textdegree{} & 25.5\textdegree{} & 29\textdegree{} \\
\hfill
\includegraphics[width=0.09\textwidth]{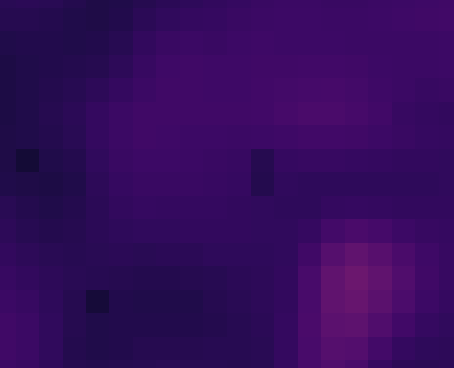} &
\includegraphics[width=0.09\textwidth]{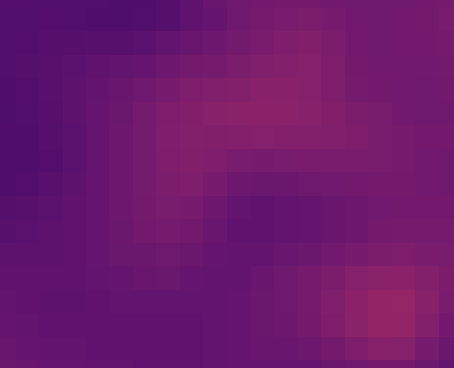} &
\includegraphics[width=0.09\textwidth]{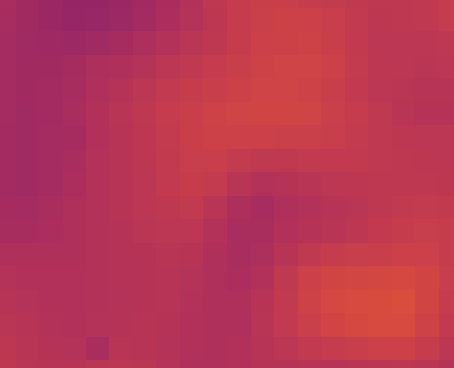} &
\includegraphics[width=0.09\textwidth]{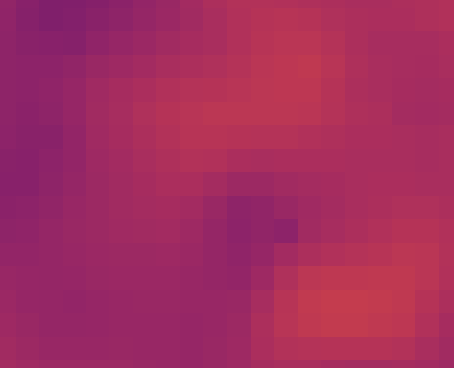} &
\includegraphics[width=0.09\textwidth]{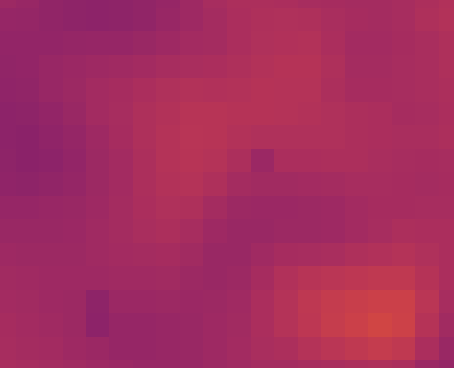} \\
\hfill
\includegraphics[width=0.09\textwidth]{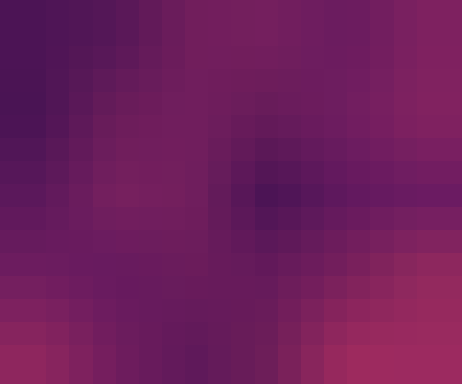} &
\includegraphics[width=0.09\textwidth]{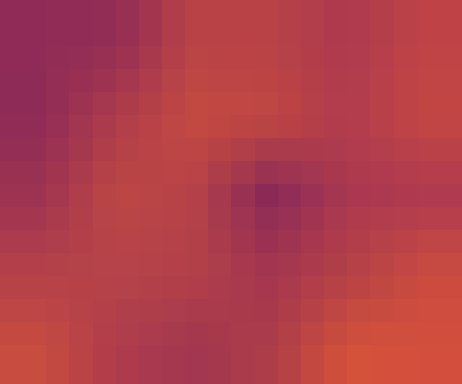} &
\includegraphics[width=0.09\textwidth]{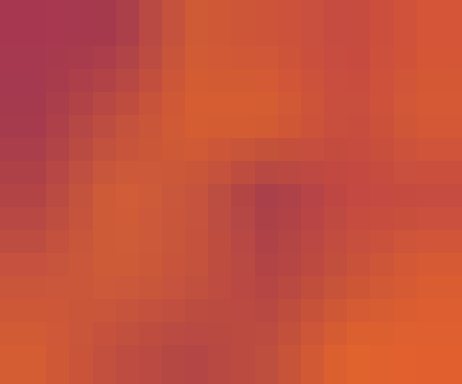} &
\includegraphics[width=0.09\textwidth]{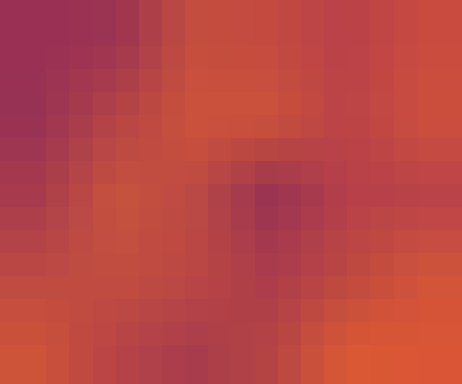} &
\includegraphics[width=0.09\textwidth]{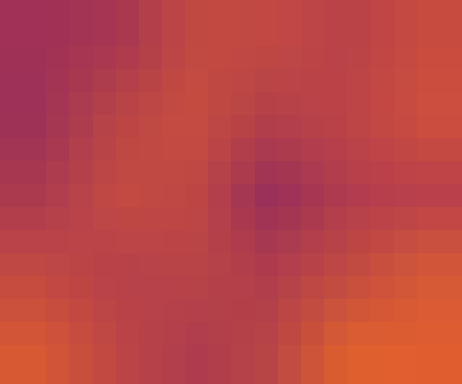} \\
\hfill
\includegraphics[width=0.09\textwidth]{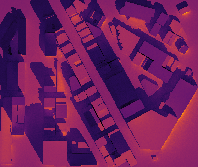} &
\includegraphics[width=0.09\textwidth]{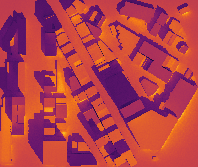} &
\includegraphics[width=0.09\textwidth]{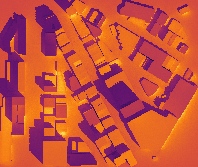} &
\includegraphics[width=0.09\textwidth]{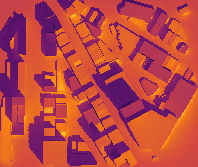} &
\includegraphics[width=0.09\textwidth]{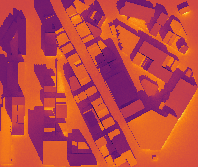} \\
\end{tabular}
\caption{\rosalieon Comparison of Landsat Surface Temperature at different dates in 2024 in Detroit (top) with our simulation (bottom), run at the corresponding date, time and air temperature. We downscale our result \revisionon{}to the same resolution as Landsat (100m/pixel) to compare (middle). The satellite records Detroit around 12:16 pm, local time. Results are all presented with the inferno range 290K-338K, as in Fig.~\ref{fig:detroit_GT_comp}. The overall gradients are very similar to Landsat ones, with a consistent shift in absolute value. \revisionoff{}\rosalieoff }

\label{fig:Detroit_Landsat_Comp}
\end{figure}

\section{Application to urban planning}

To \revisionon{}demonstrate how\revisionoff{} urban planners can benefit from this simulator, we present two studies that might be of interest for them. The first study \revisionon{}examines\revisionoff{} the temporal evolution of surface temperature over time to identify areas of heat retention (Sec.~\ref{TemporalStudy})\revisionon{}. The second study explores the impact\revisionoff{} of material changes on heat transfers (Sec.~\ref{Real city simu}). In order to highlight the impact of facade material on \revisionon{}those\revisionoff{} experiments, \revisionon{}in this section\revisionoff{} we applied \revisionon{}the same\revisionoff{} concrete to all building roofs.

\subsection{Temporal evolution study}
\label{TemporalStudy}

\revisionon{}We simulate the surface temperature evolution on June 17th, 2024, in a Detroit neighborhood described by 18 181x245 pixels maps at 0.5m/pixel resolution. The air temperature goes from 22°C (295K) up to 35°C (308K).\revisionoff{} The experimental conditions are summarized in Fig.~\ref{fig:expConditionTime}, and Fig.~\ref{fig:time_varying} shows the heatmap evolution over the day. 

\begin{figure*}[htb]   
  \centering 
    \includegraphics[width=\textwidth]{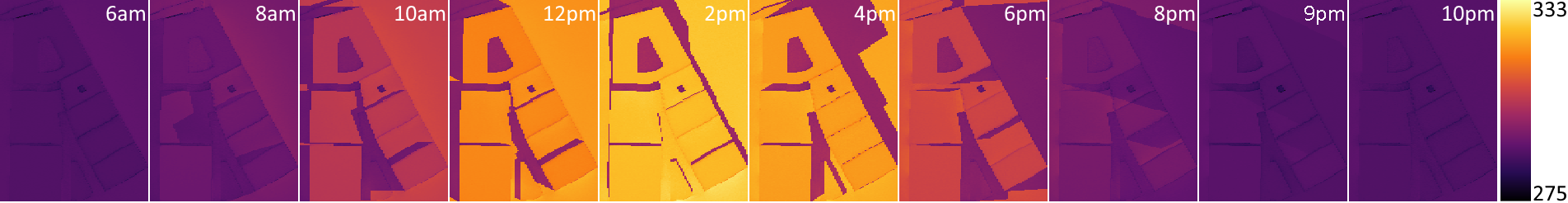}
    \caption{Temporal evolution of surface temperature over one day of one building block of Detroit.}
    \label{fig:time_varying}
\end{figure*}

 \revisionon{}Each output of Fig.~\ref{fig:time_varying} corresponds to the steady state of one simulation ahead of time, with as initial surface temperature the result of the simulation at the previous time.\revisionoff{} We notice a heat peak at 2pm, with an overall surface temperature of 330K, hotter than the air temperature, which is explained by the absorption of solar irradiation. The shadows are in accordance with the hourly sun position, and we can observe hotter areas on the ground next to the facades that receive direct solar irradiation. At 10pm, the overall temperature is the same as the air temperature, around 292K, which is also expected as the sun does not contribute anymore to surface heating. It can be noted that the surfaces at 9pm are hotter than at 6am even if the air temperature and the solar irradiation are similar, due to heat absorption by the building material. 

\begin{figure} [H]  
  \centering 
    \includegraphics[height=4cm]{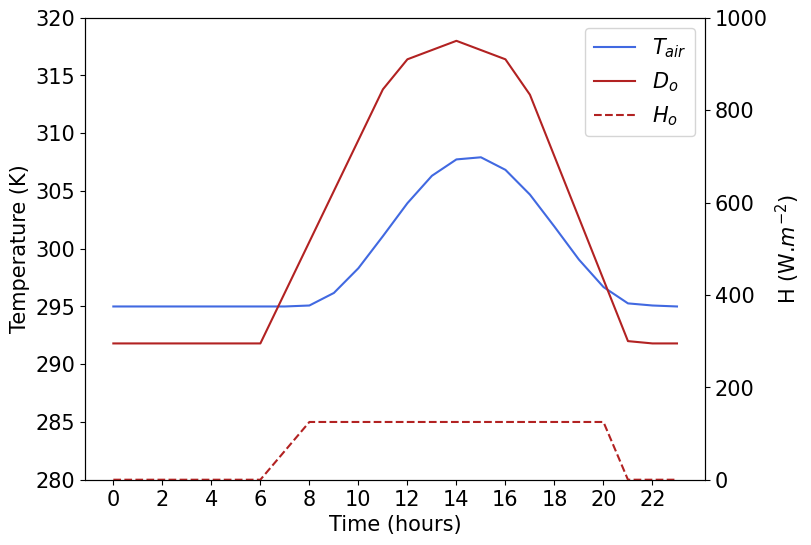}
    \caption{Experimental conditions of the time-varying air temperature and the solar irradiation.}
    \label{fig:expConditionTime}
\end{figure}

\subsection{Material change study}
\label{Real city simu}

To highlight the capability of our simulator, we investigate the influence of materials on the UHI effect \revisionon{}in a neighbourhood of the city of Detroit with a spatial resolution of 0.5m per pixel for fine-grained examination, resulting in 18 181x245 pixels maps\revisionoff{}. The simulation runs with a starting point on June 6, 2024 at noon with an assumed air temperature of 308K. The material disposition considered is the one extracted with our pipeline, before applying a change in the facade material to observe material properties' impact on the resulting surface temperatures. In the simulation shown in Fig.~\ref{fig:detroit_change_mat}, we init the simulation a) with all building facades made of concrete as the main material, keeping the same percentage estimated, and b) with limestone instead. We don't modify the roof and ground materials, that we set respectively to concrete and asphalt. We can observe in c) that limestone induced an increase in the surface temperature, up to 2.4K in some areas. Limestone has a higher emissivity than concrete (See Suppl. Tab.~\ref{table_material_database}), which implies a higher thermal absorptivity and a lower reflectivity in the visible spectrum. Thus, the limestone facades tend to irradiate more via the indirect sun path, leading to higher local temperatures near the exposed facades. Moreover, knowing that the map is north-facing, and the simulation is running for 2 pm in Detroit, the shadow positions are consistent with the sun’s position and their sizes with the district height map. 

\revisionon{}
Our method scales well to bigger area as highligted in Fig.~\ref{fig:detroit_GT_comp} which represents the Detroit district heatmap and its corresponfing heightmap around our current study area with a spatial resolution of 3m/pixel, thus covering 592x499m, but with diverse roof albedos.

\revisionoff{}


\begin{figure}[b]
  \centering 
    \includegraphics[width=0.48\textwidth]{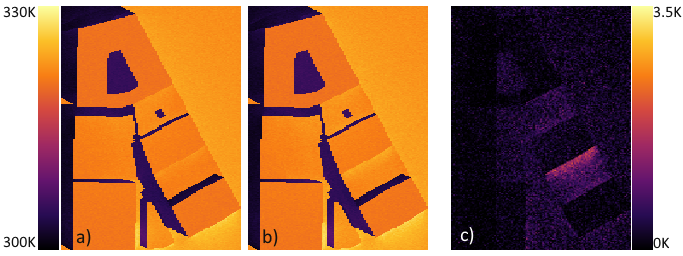}
    \caption{Simulation of a facade material change in Detroit. a) The main material of all facades is set to Concrete. b) The main material of all facades is set to Limestone. c) Difference in temperature between a) and b). With the higher emissivity of Limestone, more heat is reflected to the roofs next to a high building, with a difference of 2.4K in some areas.}
    \label{fig:detroit_change_mat}
\end{figure}



Thus, a user can, for any city where street-view images and OpenStreetMap information are available, model a neighborhood and test different material compositions to select the one with the lowest thermal impact. This could benefit urban planners to study material influence independently of other factors, such as the anthropic heat, the vegetation or the weather.
\section{Discussion and Future Work}

In this paper, we introduced a pipeline that estimates the composition of a city's building facades, coupled with 2D city features, which we encode into a set of \revisionon{}textures\revisionoff{} to feed into our 2.5D \revisionon{}coupled heat transfer\revisionoff{} simulator. This approach \revisionon{}enables the simulation of the surface temperature of an urban area, and test the impact of new material configurations on the UHI effect. The evaluation showed consistent results compared to satellite measurements, as well as to the 3D thermal simulator Stardis. Using 2.5D data instead of a 3D model enables 20$\times$ faster simulations due to the random accessibility of the maps and the reduced complexity introduced by our representative procedural facade generation. Moreover, 2D representations are an order of magnitude simpler for neural models to process than 3D ones. This paves the way for future dataset generation and learning-based approaches for UHI effect prediction.\revisionoff{}


\revisionon{} Though our approach is conceptually generalizable to any city, our solution relies heavily on the availability of open data. \revisionoff{}Consequently, the set of materials considered is limited, as literature describing both optical and thermal properties of the same material is rare. Similarly, the coverage of the Mapillary database is spatially variable as it depends on community contributions. It can be mitigated by coupling this data with other sources, like cities' own open data. Our solution depends on the VLM performance in the detection of building materials. In this study, we opted for LlaVA 1.6, but it can be substituted by any other VLM or model specifically trained for this task. Our implementation of coupled conductive transfer currently allows for heat transfer within a bounded surface, namely the facade of a building, the roof, or the ground. The transition between surfaces and the change of path direction at corners are not yet included and could be beneficial to the simulator.

In the future, we could extend our facade material estimation pipeline, by considering aerial or satellite images to automatically retrieve both roofs and ground materials. Finally, \revisionon{}to handle potential scalability issues induced by hardware specifications\revisionoff{}, our approach could benefit from a multi-resolution approach: start with a low resolution satellite land surface temperature image of a hot sunny day and determine the hottest areas of the city. Then, in those areas, run our simulation at low-resolution (3m/pixel), to identify problematic building blocks before applying our current high-resolution pipeline (0.5m/pixel) and explore different material alternatives to mitigate the local temperature increase. 



\label{Conclusion}

\bibliographystyle{eg-alpha-doi}  
\bibliography{main.bib}

\clearpage

\setcounter{section}{0}
\setcounter{figure}{0}
\setcounter{table}{0}



\section{Material estimation pipeline validation}

To measure the performance of our pipeline and validate our approach, we use multiclass IoU and accuracy metrics. As our facade element segmentation is purely metadata and not localized pixels as in a common segmentation task, we consider as true positive the common percentage of an element with the right material compared to ground truth. Each group of facade elements (ground or upper levels) must add up to 100\% in the VLM output, and we enforce it afterward if they don't, by scaling the results. Table~\ref{table_vlm_validation} presents the ablation regarding the benefits of GroundedSAM and shows performances obtained with Llav1.6~\cite{Llava1.5:2023}, InternVL2~\cite{InternVL2:2024}, and ClaudeAI~\cite{TheC3}. Even if ClaudeAI has the best results, we pursue with Llava 1.6 to use an open model in our experiments. We use the following prompt:

\begin{lstlisting}[basicstyle=\tiny]
Could you please describe the main building present in the image, 
by giving the material and proportion of each facade element ? 
Output text as a JSON. 
Separate the ground level and the upper levels. 

For the ground level, provide:
1) the main facade material name and percentage of presence, 
2) the window material name and percentage of presence,
3) the window frame material name and percentage of presence,
4) the door material name and percentage of presence. 
For the upper levels, provide:
1) the main facade material name and percentage of presence, 
2) the window material name and percentage of presence,
3) the window frame material name and percentage of presence,
4) the shutter material name and percentage of presence. 
   
Here is the list of different materials:
    Brick
    Aluminium
    Concrete
    Metal
    Glass
    Wood
    Terracotta
    Limestone
    Stone
    Cement
    Asphalt
    Slate
    
Return only a json like this:
```json
{ "ground_level": {
    "main_facade_material": "",
    "main_facade_material_percentage": "%",
    "windows_material": "",
    "windows_material_percentage": "%",
    "frames_material": "",
    "frames_material_percentage": "%",
    "doors_material": "",
    "doors_material_percentage": "%"
  },
  "upper_levels": {
    "main_facade_material": "",
    "main_facade_material_percentage": "%",
    "windows_material": "",
    "windows_material_percentage": "%",
    "frames_material": "",
    "frames_material_percentage": "%",
    "shutters_material": "",
    "shutters_material_percentage": "%"
}}
```
  \end{lstlisting}


\begin{table}[htb]
\centering
\small
\begin{tabular}{|lllll|}
\hline
Model     & Llava & Llava & InternVL2 & Claude \\
\hline
Input     & input & SAM   &   SAM     &   SAM  \\
\hline
Multi-class IoU       & 0.375 & 0.416 & 0.392 & \textbf{0.47} \\
Accuracy  & 0.521 & 0.563 & 0.521 & \textbf{0.59}\\
\hline
\end{tabular}
\caption{Material estimation performances. The first two columns demonstrate the gain in performance of GroundedSAM. The two others compare Llava with two other VLM: InternVL2 and ClaudeAI. The latter performs best.}
\label{table_vlm_validation}
\end{table}

We illustrate in Fig.\ref{fig:vlm_groundedsam} the process from the equirectangular geolocated image of Mapillary dataset (a.), that we rectify using the Perpective Fields method to get a straight building (b.), before calling GroundedSAM to get the segmentation mask containing facades, that we use to crop the image (c.). With this input, when we query Llava1.6 with our prompt, we get the following material proportion estimation:

\vspace{6pt}

\begin{lstlisting}[basicstyle=\tiny]
```json
{ "ground_level": {
    "main_facade_material": "Brick",
    "main_facade_material_percentage": "45%",
    "windows_material": "Glass",
    "windows_material_percentage": "25%",
    "frames_material": "Aluminium",
    "frames_material_percentage": "20%",
    "doors_material": "Wood",
    "doors_material_percentage": "10%"
  },
  "upper_levels": {
    "main_facade_material": "Brick",
    "main_facade_material_percentage": "45%",
    "windows_material": "Glass",
    "windows_material_percentage": "30%",
    "frames_material": "Aluminium",
    "frames_material_percentage": "20%",
    "shutters_material": "Wood",
    "shutters_material_percentage": "5%"
}}
``` 
\end{lstlisting}

\begin{figure} [H]  
  \centering 
    \includegraphics[width=0.48\textwidth]{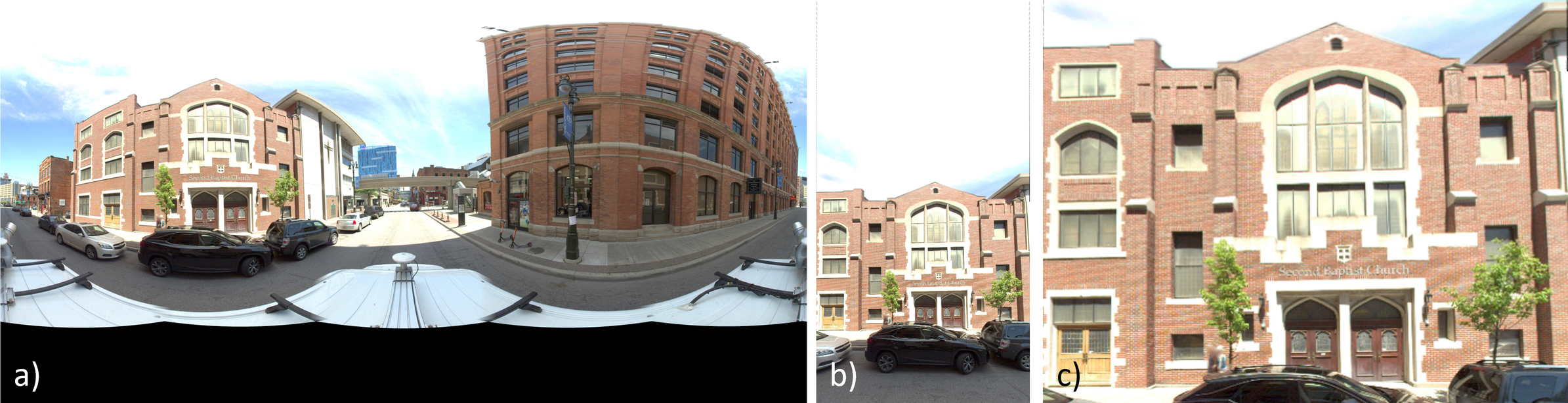}
    \caption{From 360° Mapillary image (a.) to sliced rectified facade (b.) to cropped building facade (c.).}
    \label{fig:vlm_groundedsam}
\end{figure}

\section{VLM pipeline accuracy impact on simulation results} 





\begin{table}[b]
\centering
\small
\vspace{-5pt}
\begin{tabular}{|lcccc|}
\hline
Config.      & Both wrong & Mat. only & Perc. only & Ours \\
\hline
RMSE (K)     &  2.79 & 2.72  & 2.13 & \textbf{1.92} \\
\hline
\end{tabular}
\vspace{-3pt}

\caption{Study the impact of wrong material predictions on the temperature, over 4 configurations, compared to GT.}


\label{table_study_wrong_material_impact}
\end{table}

To illustrate the impact of an inaccurate facade composition, using the same simulation parameters as in Fig. 9, we compared the ground truth material composition to four configurations: (1) both material types and percentages are incorrect, (2) wrong materials only, (3) wrong percentages only, (4) the material composition provided by our pipeline. We compared each configuration results to the ground truth using the RMSE (See Table \ref{table_study_wrong_material_impact}). Our pipeline result is accurate enough to detect material impact and capture variations, performing better than arbitrary composition. In addition, the most impactful mistake is the misdetection of material nature, rather than inaccurate percentages.

\revisionon{}
\section{Comparison to Local Climate Zone materials}

As an alternative to our VLM pipeline for material estimation, we also consider relying on coarser-grained information from existing databases.
In particular, the Local Climate Zone (LCZ) classification~\cite{LCZ:2022} characterizes urban areas as a combination of one of ten standardized build types (density, size, and types of buildings) and one of seven land cover types (soil kind and vegetation if any).
LCZ data are provided by area as an aggregate of the characteristics of buildings in that area.
In comparison, our VLM pipeline provides material estimation for each building individually, therefore offering more fine-grained modeling.

To evaluate the benefits of our VLM pipeline over LCZ data, we compare the absolute errors compared to our hand-made ground truth with both our VLM pipeline material description and the uniform one obtained with LCZ data.
Figure~\ref{fig:lcz} presents the results of the comparison, with an RMSE of 0.59 with the VLM pipeline, and 0.67 (13\% higher) when using LCZ materials.
As expected, when using LCZ data the errors close to the facades are bigger than when using data from our VLM pipeline.
Indeed, LCZ provides only aggregated data on facade materials, while our method more accurately models the specifics of each building.

\begin{figure}[htb]
  \centering 
  \begin{subfigure}[t]{0.22\textwidth}
    \includegraphics[width=\textwidth]{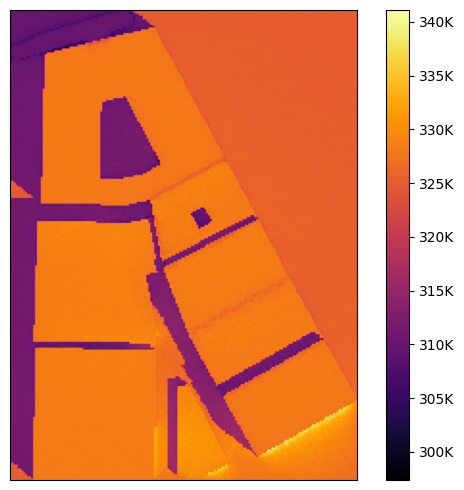}
    \caption{Ground truth materials}
  \end{subfigure}
    \hfill
  \begin{subfigure}[t]{0.21\textwidth}
    \includegraphics[width=\textwidth]{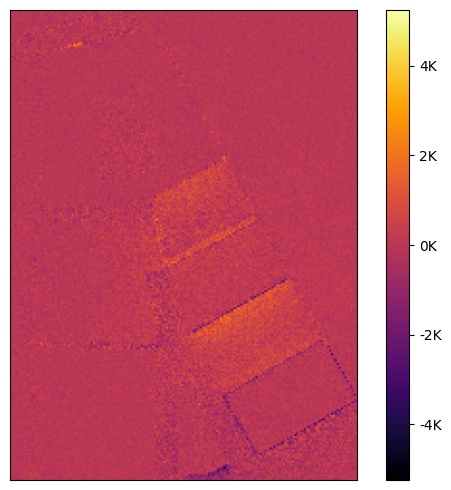}
    \caption{Difference of absolute error}
  \end{subfigure}
  \caption{
\revisionon{}Heatmap obtained with ground truth materials (a), and difference of absolute errors between heatmaps relative to the Ground Truth with LCZ and VLM materials (b).
  Positive difference means bigger error with LCZ materials, negative difference means bigger error with VLM materials.
\revisionoff{}
  }
  \label{fig:lcz}
\end{figure}

\revisionoff{}

\section{Material database}

\begin{table}[h]
\footnotesize
\begin{tabular}{|l|r|r|r|r|r|r|}
\hline
\multirow{2}{*}{Material}  & \multicolumn{1}{c|}{Th. capacity} & \multicolumn{1}{c|}{Conductivity} & \multicolumn{1}{c|}{Density} & \multirow{2}{*}{Emissivity} \\
           & \multicolumn{1}{c|}{($J/K/kg$)} & \multicolumn{1}{c|}{($W/m/K$)} & \multicolumn{1}{c|}{(kg/$m^3$)} &  \\
\hline
\rule{0pt}{2.5ex}Brick      &    790     &  0.9  &   1920    & 0.93  \\
Aluminium  &    903     &  237  &   2702    & 0.03  \\
Concrete   &    880     &  1.4  &   2300    & 0.88  \\
Steel      &    456     &  15.6 &   7913    & 0.85  \\
Glass      &    840     &  1    &   500     & 0.93  \\
Wood       &    1880    & 0.12  &   450     &  0.9   \\
Terracotta &    1800    & 0.8   &   780     &  0.6   \\
Limestone  &    1000    & 1.7   &   2200    & 0.95  \\
Stone      &    840     &  2.68 &   2550    & 0.87  \\
Cement     &    920     &  0.43 &   1283    & 0.54  \\
Asphalt    &    1000    & 0.5   &   1700    & 0.94  \\
Slate      &    1000    & 2.2   &   2400    & 0.97  \\     
\hline
\end{tabular}

\caption{Material thermal and optical properties. 
} 
\label{table_material_database}
\end{table}
The material properties considered in the experiments on the Detroit city are listed in Tab.~\ref{table_material_database}. The information has been concatenated from various data sources \cite{MaterialLitt1, MaterialLitt2, MaterialLitt3, MaterialLitt4, MaterialLitt5, MaterialLitt6, MaterialLitt7}.

\section{Common experimental settings for all experiments}

In all the experiments mentioned in the context of the paper, the results presented correspond to about 5000 paths initiated per pixel. When involved, the radiative and indirect irradiation paths can bounce at most 30 times. For the conductive path, 700 steps are performed before switching to another heat transfer mode. \revisionon{}The number of steps is set so that enough of the facade space is explored to reach the initial time to account for conductive transfer in the estimated temperature. It is set based on empirical observations against the reference provided by satellite imagery for a short time period of a few hours.\revisionoff{} We perform at most 100 transitions between heat transfer modes if the path doesn't reach an end condition. The path is thus considered invalid and is not involved in the average computation. 

\section{Stardis experimental settings}
\begin{figure}[htb]
  \centering 
    \includegraphics[width=0.48\textwidth]{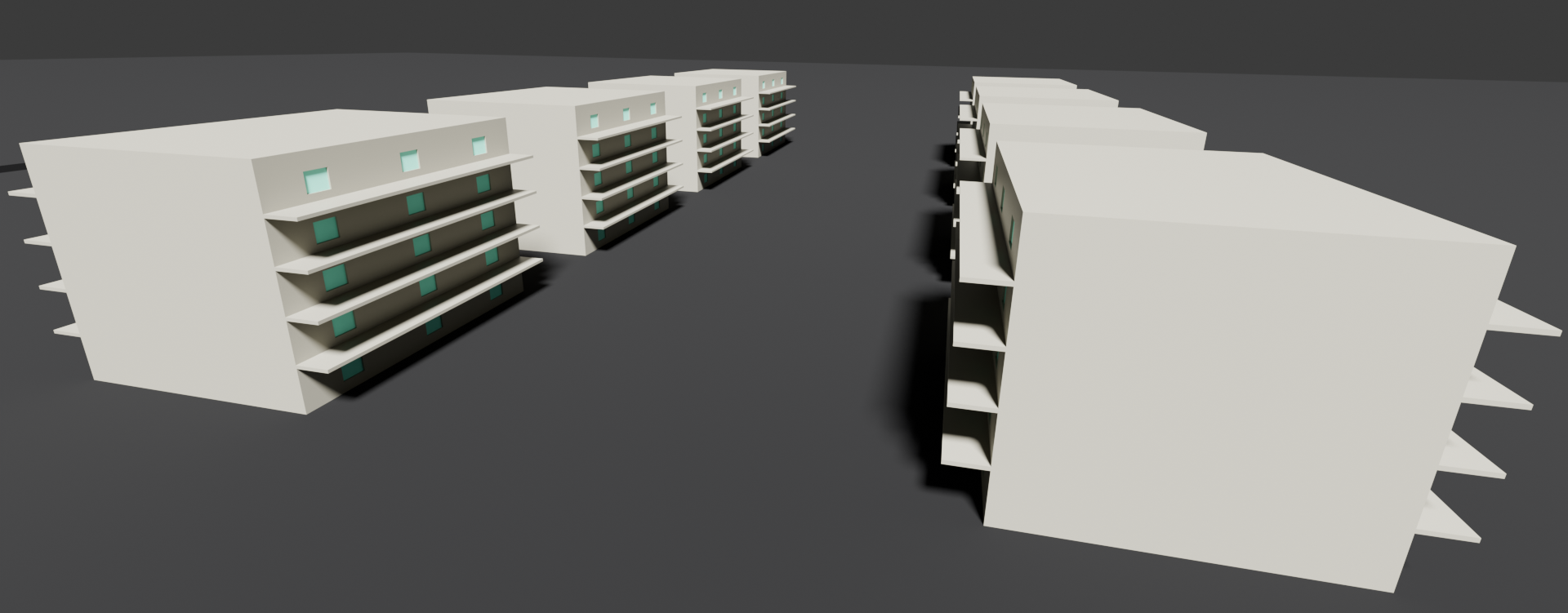}
    \caption{3D geometry input of Stardis, converted into a set of 2D maps for comparison.}
    \label{fig:ValGeometry}
\end{figure}

For the comparison with the Stardis simulator, we consider the simple city geometry provided by the Stardis Starter Pack~\cite{StardisStarterPack} without the balconies and the lake elements, as shown in Fig.~\ref{fig:ValGeometry}. It is composed of eight buildings 20m long, 15m large, and 13.7m high. Buildings of a row are separated by 10m, and the two rows by 35m. The initial conditions of each material in the model are described by Tab.~\ref{table_material_comp_stardis}. In the context of our simulator, we set the facade sampling parameters, extracted from the 3D model, gathered in Tab.~\ref{table_facade_comp_stardis}. In addition, the air temperature is set to 300K and the sky temperature to 280K.

\setlength{\tabcolsep}{0.2em}
\begin{table}[h]
\footnotesize
\begin{tabular}{|l|r|r|r|r|r|r|}
\hline
\multirow{2}{*}{Material}  & \multicolumn{1}{c|}{Th. capacity} & \multicolumn{1}{c|}{Conductivity} & \multicolumn{1}{c|}{Density} & \multirow{2}{*}{Emissivity} & \multicolumn{1}{c|}{Initial Temp. }  \\
           & \multicolumn{1}{c|}{($J/K/kg$)} & \multicolumn{1}{c|}{($W/m/K$)} & \multicolumn{1}{c|}{(kg/$m^3$)} & & \multicolumn{1}{c|}{(K)} \\
\hline
\rule{0pt}{2.5ex}Wall  &    2000     &  1.5  &   2500    & 0.9& 273 \\
Glazing   &    2000     &  0.9  &   2500    & 0.9 & 273 \\
Ground      &    2000     &  1 &   2500    & 0.9 & 273 \\
 
\hline
\end{tabular}

\caption{Material thermal and optical properties and their initial temperature as input of our simulator for the comparison to Stardis. 
} 
\label{table_material_comp_stardis}
\end{table}
\setlength{\tabcolsep}{0.4em}
\begin{table}[!h]
\footnotesize
\centering
\begin{tabular}{|l|r|r|r|r|r|r|r|r|r|}
\hline
\rule{0pt}{2ex} Pattern Width (m) & 6.6\\

\rule{0pt}{2ex} Pattern Height (m) & 2.7\\ 

\rule{0pt}{2ex} Max door width (m) & 4\\

\rule{0pt}{2ex} Facade perimeter (m) & 68.6\\ 

\rule{0pt}{2ex} Facade height (m) & 13.7\\ 

\rule{0pt}{2ex} Upper Floors Window Coverage (\%) & 6  \\ 

\rule{0pt}{2ex} Ground Floor Window Coverage (\%) & 6 \\ 

\rule{0pt}{2ex} Upper Floors Facade Coverage (\%)  & 94 \\ 

\rule{0pt}{2ex} Ground Floor Facade Coverage (\%) & 94 \\

\rule{0pt}{2ex} Ground Floor Door Coverage (\%) & 0\\ 

\hline
\end{tabular}

\caption{Procedural facade sampling parameters used for the comparison to Stardis. 
} 
\label{table_facade_comp_stardis}
\end{table}

\revisionon{}
\section{Study of the convergence of our simulator }

To study the convergence of our simulator, we compute the same simulation with an increasing number of pixels and compute the FLIP \cite{Flip:2020} error against the image obtained after the equivalent of 15000 samples per pixel in Stardis. The FLIP indicator enables to quantify the differences between rendered images and corresponding ground truths, which corresponds to the image assumed converged in our case. Figure~\ref{fig:FLIPerror} demonstrates that after 5000 samples per pixels, the difference between two rendered images decreases slowly, which can constitute a sweet spot for convergence.
\begin{figure}[htb]
  \centering 
    \includegraphics[width=0.37\textwidth]{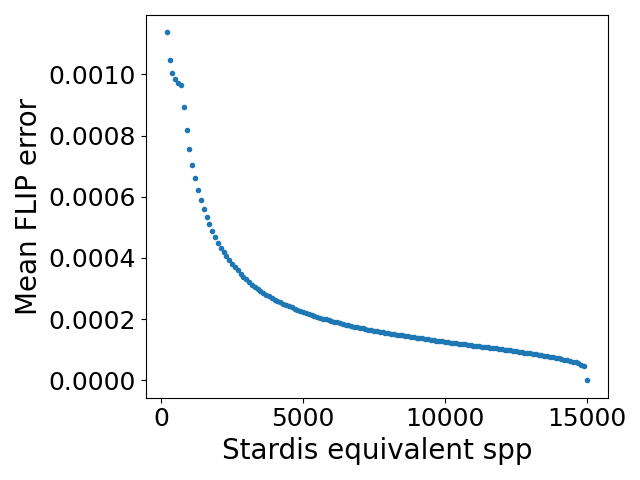}
    \caption{\revisionon{}Evolution of the FLIP value according to the number of sample per pixels equivalent in Stardis simulation.\revisionoff{}}
    \label{fig:FLIPerror}
\end{figure}

\revisionoff{}

\section{The colorized procedural facade generation}

\begin{figure*}[h]  
  \centering 
    \includegraphics[width=0.9\textwidth]{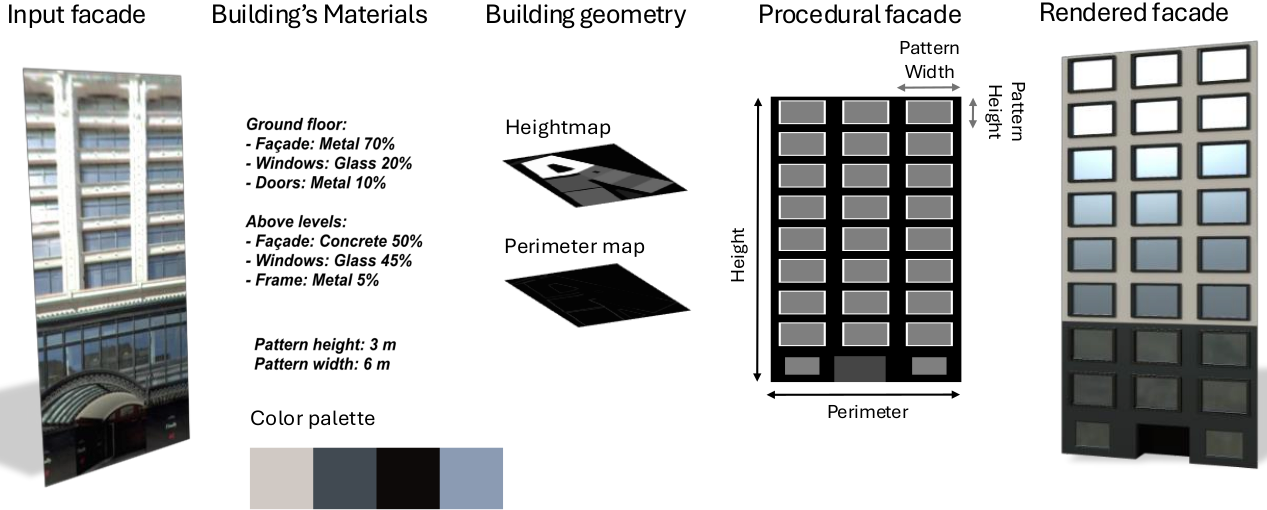}
    \caption{A full pipeline providing a realistic rendering of the facade by also extracting the color palette from the street-view image. Realism can be further improved by combining it with procedural material.}
    \label{fig:RealFacade}
\end{figure*}

Figure~\ref{fig:RealFacade} details the process of generating the procedural facade and introduces how it can produce a more realistic rendering using our pipeline. As already described in the paper, the procedural facade is generated based on the facade composition extracted from the street-view image of the facade using a Vision Language Model (VLM). In addition, it also relies on the facade perimeter and the building height in order to create a facade pattern that could be superimposed on extruded 3D geometry like wallpaper. The facade unit element pattern to be repeated depends on two extra parameters. The pattern width, meaning the portion of the wall in which we expect to have one window with its surrounding elements like a frame or shutters horizontally. The pattern height corresponds to the floor height. These values are constants and specified for each city according to its main features. In addition, for a more realistic rendering, the segmented main facade extracted by the Segment Anything Model is given as input of a color palette extractor, thus providing the set of the 4th majoritarian colors in the facades as well as their occurrence.

By combining both the procedural facade and the color palette, we can procedurally generate a simplified version of the real facade statistically coherent, to be further rendered with the right colors. It could be even a basis to create a realistic view by applying each material its own procedural textures, parametrized by the extracted colors, which could also embed architectural features specific to the studied city.


\end{document}